\documentclass[10pt, preprint]{emulateapj}

 \slugcomment{accepted by ApJ Supplement}

 \shorttitle{Underflight calibration of SOHO/CDS and Hinode/EIS with EUNIS-07}
 \shortauthors{Wang et al.}

\begin{document}

 \title{Underflight calibration of SOHO/CDS and Hinode/EIS with EUNIS-07}

 \author{Tongjiang Wang\altaffilmark{1,2}, Roger J. Thomas\altaffilmark{2}, 
  Jeffrey W. Brosius\altaffilmark{1,2}, Peter R. Young\altaffilmark{3}, 
  Douglas M. Rabin\altaffilmark{2}, Joseph M. Davila\altaffilmark{2}, 
  Giulio Del Zanna\altaffilmark{4}}

 \altaffiltext{1}{Institute for Astrophysics and Computational Sciences (IACS) 
   in the Department of Physics, Catholic University of America, 
   620 Michigan Avenue NE, Washington, DC 20064, USA; tongjiang.wang@nasa.gov}
 \altaffiltext{2}{NASA Goddard Space Flight Center, Code 671, Greenbelt, MD 20771, USA}
 \altaffiltext{3}{College of Science, George Mason University, 4400 University Drive, 
    Fairfax, VA 22030, USA}
 \altaffiltext{4}{DAMTP, Centre for Mathematical Sciences, University of Cambridge, 
  Wilberforce Road, Cambridge, CB3 0WA, UK}

\begin{abstract}
Flights of Goddard Space Flight Center's Extreme-Ultraviolet Normal-Incidence
Spectrograph (EUNIS) sounding rocket in 2006 and 2007 provided
updated radiometric calibrations for SOHO/CDS and Hinode/EIS. EUNIS
carried two independent imaging spectrographs covering wavebands of 300$-$370~\AA\ 
in first order and 170$-$205~\AA\ in second order. After each
flight, end-to-end radiometric calibrations of the rocket payload
were carried out in the same facility used for pre-launch calibrations of
CDS and EIS. During the 2007 flight, EUNIS, SOHO CDS and Hinode EIS observed
the same solar locations, allowing the EUNIS calibrations to be directly
applied to both CDS and EIS. The measured CDS NIS 1 line intensities 
calibrated with the standard (version 4) responsivities with the standard 
long-term corrections are found to be too low by a factor of 
1.5 due to the decrease in responsivity. The EIS calibration update is 
performed in two ways. One is using the direct calibration transfer of 
the calibrated EUNIS-07 short wavelength (SW) channel. The other is using 
the insensitive line pairs, in which one member was observed
by EUNIS-07 long wavelength (LW) channel and the other by EIS in either 
LW or SW waveband. Measurements from both methods are in good agreement, 
and confirm (within the measurement uncertainties) the EIS responsivity 
measured directly before the instrument's launch. The measurements also 
suggest that the EIS responsivity decreased by a factor of about 1.2 
after the first year of operation (although the size of the measurement 
uncertainties is comparable to this decrease). The shape of the 
EIS SW response curve obtained by EUNIS-07 is consistent with the one
measured in laboratory prior to launch. The absolute value of the quiet-Sun
He\,{\sc{ii}} 304 \AA\  intensity measured by EUNIS-07 is consistent 
with the radiance measured by CDS NIS in quiet regions near the disk center 
and the solar minimum irradiance obtained by CDS NIS and SDO/EVE recently.
\end{abstract}


\section{Introduction}
The extreme ultraviolet (EUV; 150 -- 1200 \AA) waveband is replete with emission
lines formed at temperatures that range from several times $10^4$ K to several
times $10^7$ K, and is thus well suited for studying numerous features and
phenomena that occur in the solar atmosphere.  Like most solar EUV instruments,
the Solar EUV Research Telescope and Spectrograph (SERTS) sounding rocket
payload observed only part of the EUV waveband.  SERTS was designed to improve
on earlier solar EUV instrumentation by (1) retaining the stigmatic imaging
capability of the NRL spectroheliograph on {\em Skylab} \citep{tou77} but
with a spatial selection capability that reduced image overlap, and (2)
achieving high spectral resolution.  The version of SERTS flown in 1989
\citep[SERTS-89;][]{neu92} gathered EUV radiation with a grazing incidence
telescope (as did subsequent SERTS flights), and carried a standard gold-coated
toroidal diffraction grating.  SERTS-89 observed hundreds of first-order
emission lines between 235 and 450 \AA, as well as dozens of second-order lines
between 170 and 225 \AA\  \citep{tho94, you98}.
The version of SERTS flown in 1991 \citep[SERTS-91;][]{bro93, bro96, fal94} 
and 1993 \citep[SERTS-93;][]{bro96, bro97a, bro97b}
incorporated a multilayer-coated diffraction grating that enhanced the
instrument's first-order responsivity over that of the gold-coated grating by
factors up to nine.  The version flown in 1995 
\citep[SERTS-95;][]{bro98a, bro98b, bro99} incorporated a
multilayer-coated grating that enhanced the instrument's second-order ($\sim$
170 -- 225 \AA) responsivity, thus bringing out many lines that had not been seen
during any of the previous flights, and allowing them to be measured with the
highest spectral resolution ($\sim$ 30 m\AA) ever achieved for spatially
resolved active region and quiet-Sun spectra in this wavelength range.

 \begin{figure*}
 \epsscale{1.0}
 \plotone{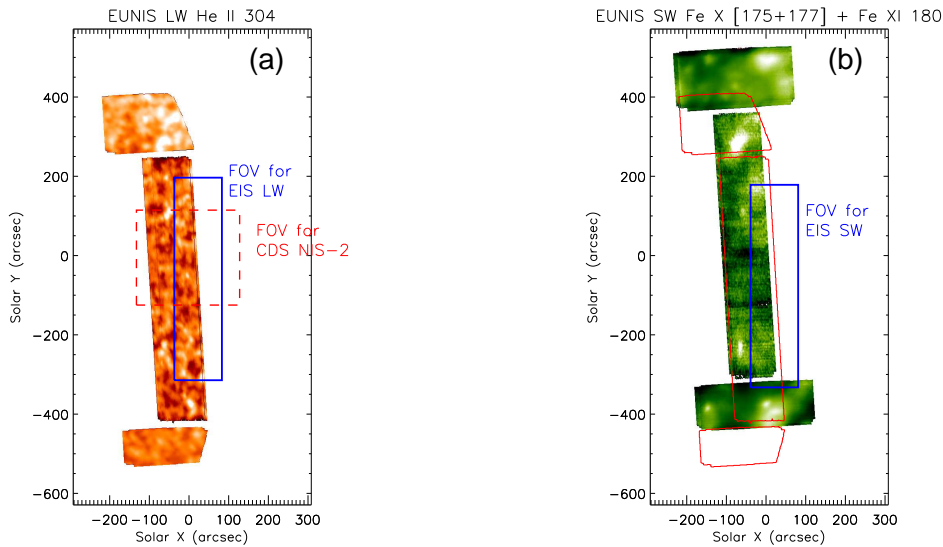}
 \caption{ \label{fgfov}
 EUNIS-07 composite full-raster images for LW and SW channels, observed from 18:02 
to 18:07 UT on 2007 November 6. (a) The LW raster image of the He\,{\sc{ii}} 304 \AA\
 line. (b) The SW image made of sums of Fe\,{\sc{x}} 174.5 \AA, Fe\,{\sc{x}} 177.2\AA, 
and Fe\,{\sc{xi}} 180.4 \AA\ lines. The dashed box in (a) indicates the FOV for 
CDS NIS-2, to which the pointing of CDS NIS-1 is about 9$^{''}$ northward shifted. 
The solid boxes in (a) and (b) indicate the FOVs for Hinode/EIS LW and SW channels,
 respectively. The overlaid contours (thin line) in (b) indicate the position of 
the LW raster image.}
 \end{figure*}

For the SERTS flights of 1989--1995 data were recorded on EUV-sensitive
photographic film that was developed and digitized after each flight.  For the
flights of 1989--1993 the relative radiometric calibration was derived by
combining laboratory measurements of component responsivities with ``insensitive
ratios" (solar emission line pairs whose intensity ratios are insensitive to
variations in the electron density and temperature), while for the flight of
1995 the relative radiometric calibration was derived strictly by the
insensitive ratio method.  In all four cases the relative radiometric
calibration was placed onto an absolute scale by requiring the sum of the He\,{\sc{ii}}
303.78 \AA\ and Si\,{\sc{xi}} 303.32 \AA\ quiet-Sun intensities to match the values
reported by \citet{man78} based on earlier observations from satellites
and sounding rockets.  In \citet{man78} the summed intensity is 7115 ergs
cm$^{-2}$s$^{-1}$sr$^{-1}$ at disk center and brightens by about 30\% toward
the limb.  The absolute radiometric calibration of SERTS derived this way was
estimated to be accurate within a factor better than 2.

The version of SERTS flown in 1997 \citep[SERTS-97;][]{bro00a,bro00b} 
incorporated the same multilayer-coated toroidal diffraction grating that was flown
in 1991 and 1993, but recorded spectrographic data on an intensified CCD-detector.
Its spectral bandpass was 299 -- 353 \AA, with an instrumental resolution (FWHM) of
115 m\AA.  An end-to-end radiometric calibration of SERTS-97 was performed after the
flight at Rutherford-Appleton Laboratory (RAL), UK, in the same facility used to 
calibrate the Coronal Diagnostic Spectrometer \citep[CDS;][]{har95} on the {\em Solar
and Heliospheric Observatory} ({\em SOHO}) spacecraft and the Extreme-ultraviolet
Imaging Spectrometer \citep[EIS;][]{cul07} aboard the {\em Hinode} satellite 
\citep{kos07}, and using the same EUV light source re-calibrated by 
Physikalisch-Technische Bundesanstalt (PTB) against the synchrotron radiation 
of BESSY I (a primary radiation standard).  
This was the first time that the SERTS instrument underwent an
absolute radiometric calibration while fully assembled.  The uncertainty on the
SERTS-97 absolute radiometric calibration was 15\%.  Coordinated, cospatial,
time-invariant SERTS-97 and CDS spectra were used to carry out an intensity
cross-calibration that yielded an improved responsivity curve for the CDS Normal
Incidence Spectrometer's NIS 1 (308 -- 380 \AA) waveband \citep{tho99, tho02}.  
The importance of a reliable radiometric calibration for SERTS,
CDS, EIS, or any EUV instrument cannot be over emphasized.  Without it, quantitative
analyses of properties like temperature, density, emission measure, and element abundances, along
with the physical interpretations that emerge from studies of those properties, are
impossible.

 \begin{figure*}
 \epsscale{1.0}
 \plotone{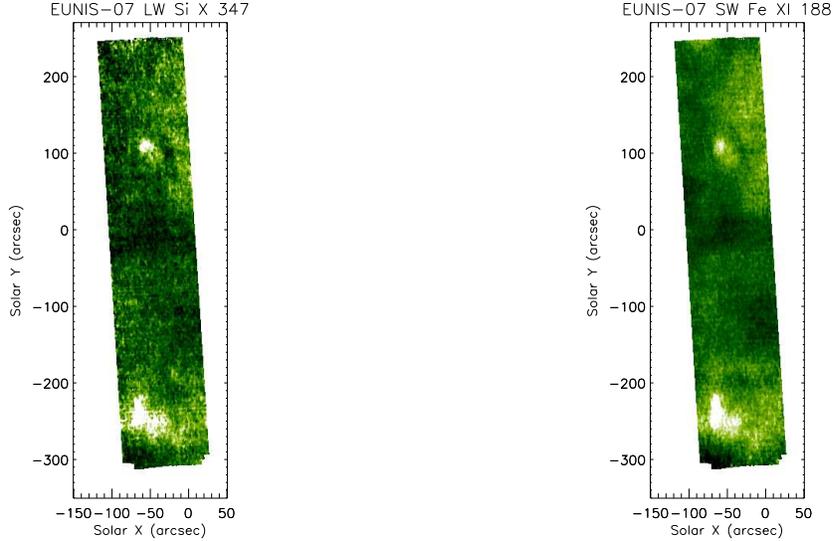}
 \caption{ \label{fgefv}
Common field of view between the slit raster images for EUNIS-07 LW and SW channels. 
(a) The LW raster image in Si\,{\sc{x}} 347.4 \AA. (b) The SW raster image in 
Fe\,{\sc{xi}} 188 (188.2$+$188.3) \AA. }
 \end{figure*}

The Extreme Ultraviolet Normal Incidence Spectrograph (EUNIS) is the successor to
SERTS.  EUNIS contains two independent but co-pointing spectrographs (each with a
normal incidence telescope), one of which covers a short wavelength (SW; 170--205
\AA) channel while the other covers a long wavelength (LW; 300--370 \AA) channel.
The instrument and its end-to-end absolute radiometric calibration is described
below.  Spectra obtained from the 2006 flight (EUNIS-06) were used to investigate
a bright point \citep{bro07, bro08b}, a cool transient brightening \citep{bro08a}, 
and transition region velocity oscillations \citep{jes08}, as
well as to derive a calibration update for CDS NIS 1 \citep{wan10}.  EUNIS was
last flown on 2007 November 6, when the solar disk contained no active regions.
EUNIS-07 obtained coordinated observations of quiet-Sun areas near disk center with
both CDS and EIS.  In what follows we derive updates to the CDS and EIS absolute
radiometric calibrations based on these coordinated observations.  Section 2
describes the observations and data reduction; \S 3 discusses the absolute radiometric
calibration of the EUNIS-06 and EUNIS-07 long wavelength channel; in \S 4 we derive
the calibration update for CDS NIS with EUNIS-07; \S 5 presents the absolute radiometric
calibration of the EUNIS-07 SW channel; in \S 6 we derive the calibration update for
both EIS channels with EUNIS-07; and in \S 7 we discuss and summarize our
conclusions.
 
 \begin{figure*}
 \epsscale{1.0}
 \plotone{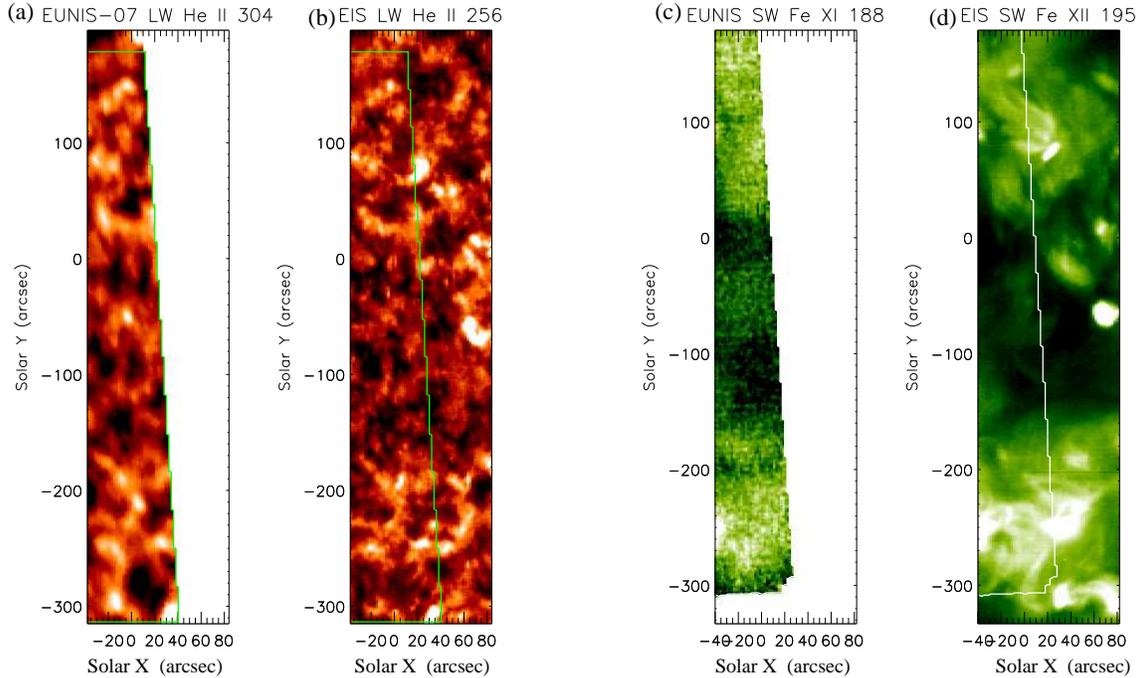}
 \caption{ \label{fgcfv}
Coalignments of EUNIS-07 and Hinode/EIS images. (a) The EUNIS-07 LW He\,{\sc{ii}} 
304 \AA\ raster image in the FOV of EIS LW channel. (b) The EIS LW He\,{\sc{ii}} 
256 \AA\ raster image observed from 18:02 to 18:54 UT. (c) The EUNIS-07 SW 
Fe\,{\sc{xi}} 188 (188.23$+$188.29) \AA\ raster image in the FOV of EIS SW channel. 
(d) The EIS SW Fe\,{\sc{xii}} 195 \AA\ raster image observed from 18:02 to 18:54 UT.
 The contour in (a) and (b) represents the common FOV between EUNIS-07 LW and EIS 
LW and SW channels. The contour in (d) represents the common FOV between
EUNIS-07 SW and EIS SW channels. }
 \end{figure*}

\section{Observations and data reduction}
EUNIS-07 was launched from White Sands Missile Range, New Mexico, at 18:00 UT on 
2007 November 6. It has a payload mass of 473 kg and reached an apogee of 303 km, 
spending 369 s above 150 km where low EUV extinction permits science-quality data. 
A total of 257 exposures, each containing solar spectra and images, 
were recorded between 1802 and 1808 UT 
with a frame rate of 1.3 s. EUNIS observed the quiet region with bright points near 
the solar disk center. Coordinated observations were obtained with the {\em Hinode}/EIS, 
the EUV Imaging Telescope \citep[EIT;][]{del95} and the CDS aboard 
{\em SOHO}, as well as the {\em Transition Region and Coronal Explorer} 
\citep[$TRACE$;][]{han99}. We coalign the data from different instruments using 
a cross-correlation method by taking EIT images as reference.

 \begin{figure*}
 \epsscale{1.0}
 \plotone{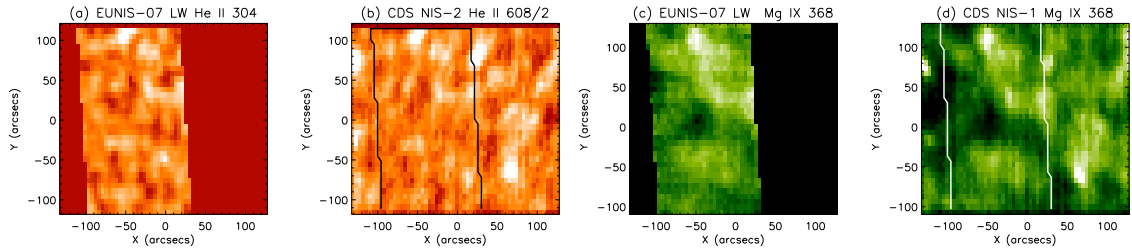}
 \caption{ \label{fgcdsmap}
 Common fields of view between the slit raster images for EUNIS-07 LW channel and 
CDS NIS. (a) The EUNIS-07 LW raster image in He\,{\sc{ii}} 304 \AA. (b) The CDS NIS-2
raster image in He\,{\sc{ii}} 608/2 \AA\ (2nd order), observed from 17:00 to 19:01 UT.  
(c) The EUNIS-07 LW raster image in Mg\,{\sc{ix}} 368 \AA. (d) The CDS NIS-1 raster 
image in Mg\,{\sc{ix}} 368 \AA. The contours in (b) and (d) represent the common 
FOV between EUNIS-07 LW and CDS NIS-2/NIS-1, over which the average
spectra are obtained for cross-calibration.}
 \end{figure*}

\subsection{EUNIS}
EUNIS comprises two independent, spatially co-aligned telescope/spectrographs 
of identical optical design, one covering EUV lines between 300 and 370 \AA\ 
seen in first order (LW channel), and a second covering lines between 170 and 205 \AA\ 
seen in second order (SW channel). Each telescope is a 110-mm diameter off-axis 
parabola that forms a real image on a precision slit formed from
a single-crystal silicon wafer using a technique developed at GSFC. The slit image 
is dispersed, magnified and reimaged by a toroidal grating onto the entrance face of 
a microchannel plate (MCP) intensifier. Each slit has a ``dumbbell" configuration. 
The center 660$^{''}$ is a conventional narrow slit (2$^{''}$). Above and below the
narrow slit is a 150$^{''}\times$200$^{''}$ ``lobe" that acts as a slitless 
spectrograph to produce monochromatic context images in strong emission lines. 
These images have proven to be extremely useful for coaligning EUNIS spectra with 
other instruments. EUNIS has a designed first-order spectral dispersion of 
25 m\AA\ pixel$^{-1}$ and a designed spatial scale of 0$^{''}$.927 pixel$^{-1}$ 
\citep{tho01}. For the flight of EUNIS-06, its optics limited the actual spatial 
resolution to about 5$^{''}$ and the measured spectral resolution was $\sim$200 
and $\sim$100 m\AA\ FWHM in the LW and SW channels.

For the flight of EUNIS-07, the lobe-slit fields of view (FOVs) were oriented 
North-South near the center of the solar disk. In this flight all observations 
were made in scanning mode. The solar image was scanned perpendicular 
to the slit at a continuous rate of about 2.4$^{''}$~s$^{-1}$ with full spectral 
images on all detectors recorded every 1.3 s (maximum frame rate). Pointing began at 
nominal field center, scanned eastward to $-$50$^{''}$, reversed, scanned westward to 
$+$50$^{''}$, and returned to center. This operation was repeated until the door was
closed, giving almost 7 complete 660$^{''}\times$100$^{''}$ spectroheliograms 
in addition to 253 lobe images in LW channel and 256 in SW channel. 
All of the raw data were processed 
with several routine adjustments, including dark image subtraction, flat-fielding 
and non-linearity correction before being converted from the recorded Data Numbers (DN) 
into Relative Exposure Units (REU), which are then used in all subsequent analyses.

\subsection{Hinode/EIS}
EIS has both imaging (40$^{''}$ and 266$^{''}$ slots) and spectroscopic (1$^{''}$ 
and 2$^{''}$ slits) capabilities, in the short wavelength (SW) range of 170$-$210 \AA\
and the long wavelength (LW) range of 250$-$290 \AA. EIS has a spectral 
resolution of about 55 m\AA\ and a spatial resolution of about 3--4$^{''}$ per pixel. 
Its spectroscopic mode can operate in a rastering mode (repeated exposures 
while scanning over the observation target) or 
a sit-and-stare mode (repeated exposures at the same spatial location). 
The {\em Hinode} spacecraft tracks the solar rotation.

The EIS observations were conducted on 2007 November 6 using EIS Study 209 
{\it EUNIS\_EIS\_Cross\_Calibration}. Two rastered images were obtained 
using 2$^{''}$$\times$512$^{''}$ slit at 60 or 61 positions with 50 s exposures. 
One began at 17:09:41 and ended at 18:00:43 UT prior to the EUNIS flight, and 
the other from 18:02:41 to 18:54:34 UT started during the EUNIS flight.  
The EIS sequence includes 5 spectral windows in SW covering nearly the full spectrum and
17 windows in LW covering selected spectral lines. The raw data were processed by 
the standard routine {\em eis\_prep} provided by SolarSoftWare (SSW) to 
remove detector bias and dark current, hot pixels, and cosmic rays, 
and to make absolute radiometric calibration. The EIS slit tilt and orbital 
variation in the line centroids were also removed from the data. 
The pointing between EIS SW and LW detectors has offsets of 2$^{''}$ in the x-direction
and about 18$^{''}$ in the y-direction \citep{you07}. 

 \begin{figure}
 \epsscale{1.0}
 \plotone{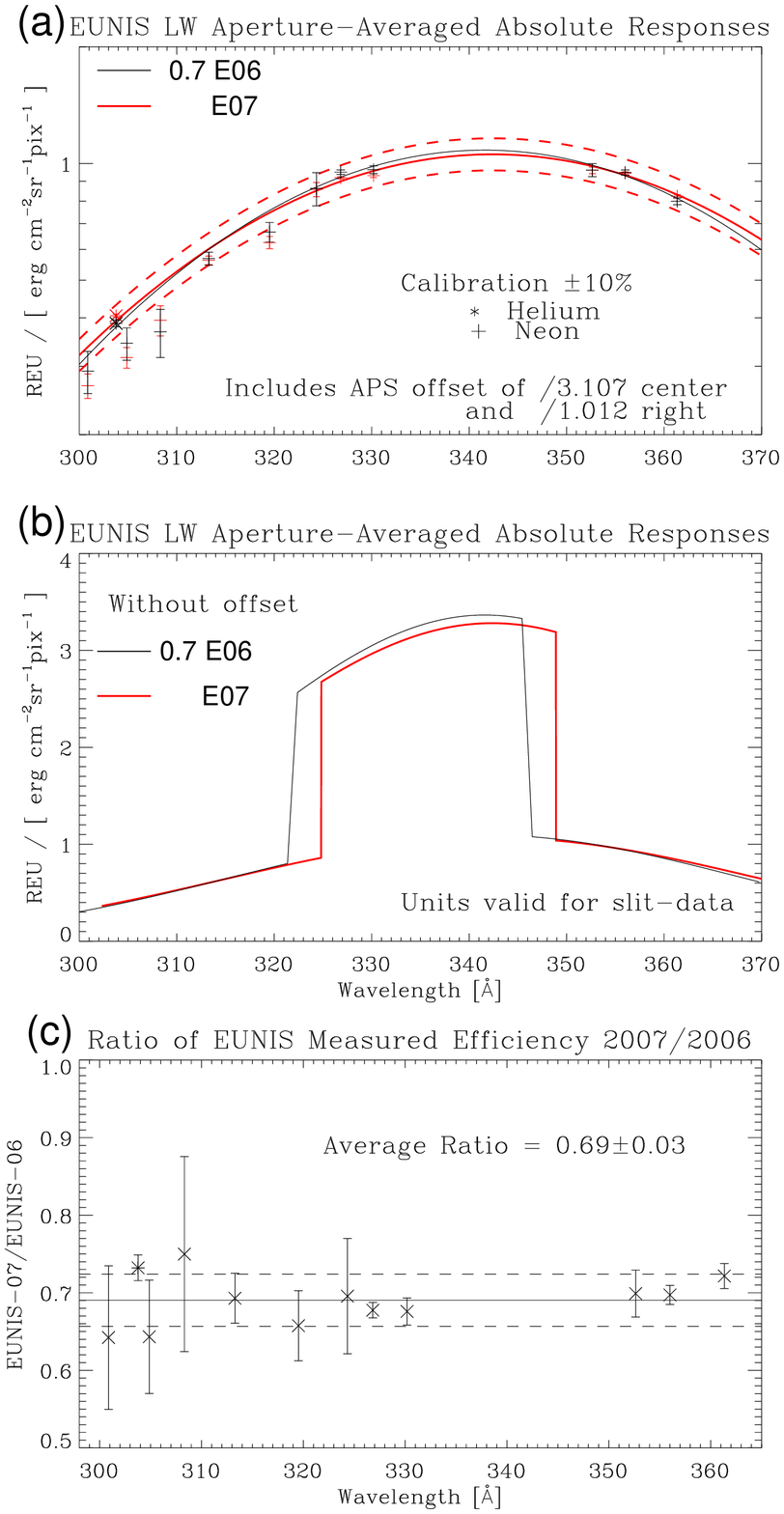}
 \caption{ \label{fglwc}
    Comparison between EUNIS-07 and EUNIS-06 LW channel calibrations. (a) Measured
instrument responsivity using the standard light source at RAL. The solid thick 
(for EUNIS-07) and thin (for EUNIS-06, reduced by a factor of 0.7) curves are 
a least-squares parabolic fit to the data. The dashed curves are the 10\% uncertainty 
for EUNIS-07. (b) The same calibration curves as in (a) but on a linear scale
and without correction of relative sensitivity factors for the detectors. 
(c) Ratio of measured responsivities for EUNIS-07 to EUNIS-06 LW channel. 
The solid line represents the average and the dashed lines its standard deviation.} 
 \end{figure}

\begin{deluxetable*}{llccccccc}
 \tabletypesize{\scriptsize}
 \tablecaption{ Quiet Region Line intensities (erg s$^{-1}$cm$^{-2}$sr$^{-1}$) for EUNIS-07 LW and 
CDS\tablenotemark{a} 
\label{tabcds}}
 \tablewidth{0pt}
 \tablehead{
 \colhead{Wavelength(\AA)} & \colhead{Ion} & \colhead{I$_{e07}$} & \colhead{I$_{CDS}^{SN}$} & \colhead{I$_{CDS}^{GZ}$} & \colhead{I$_{CDS}^S$} & \colhead{I$_{e07}$/I$_{CDS}^{SN}$} & \colhead{I$_{e07}$/I$_{CDS}^{GZ}$} & \colhead{I$_{e07}$/I$_{CDS}^S$}}
 \startdata
 303.78  &   He\,{\sc{ii}}  &   4759  &   5244  &   4484  &   4290  & 0.91 $\pm$ 0.13  & 1.06 $\pm$ 0.15  & 1.11 $\pm$ 0.16  \\
 313.76  & Mg\,{\sc{viii}}  &   25.2  &   31.7  &   25.9  &   14.1  & 0.79 $\pm$ 0.11  & 0.97 $\pm$ 0.14  & 1.79 $\pm$ 0.25  \\
 314.31  & Si\,{\sc{viii}}  &   25.1  &   36.1  &   30.0  &   24.1  & 0.70 $\pm$ 0.10  & 0.84 $\pm$ 0.12  & 1.04 $\pm$ 0.15  \\
 315.01  & Mg\,{\sc{viii}}  &   62.3  &   37.6  &   31.7  &   22.0  & 1.66 $\pm$ 0.23  & 1.97 $\pm$ 0.28  & 2.83 $\pm$ 0.40  \\
 316.20  & Si\,{\sc{viii}}  &   39.2  &   44.7  &   38.7  &   26.6  & 0.88 $\pm$ 0.12  & 1.01 $\pm$ 0.14  & 1.47 $\pm$ 0.21  \\
 319.81  & Si\,{\sc{viii}}  &   60.5  &   48.6  &   45.5  &   26.4  & 1.24 $\pm$ 0.18  & 1.33 $\pm$ 0.19  & 2.29 $\pm$ 0.32  \\
 341.91  &   Si\,{\sc{ix}}  &   15.0  &   17.0  &   15.4  &   13.9  & 0.88 $\pm$ 0.12  & 0.97 $\pm$ 0.14  & 1.08 $\pm$ 0.15  \\
 345.04  &   Si\,{\sc{ix}}  &   34.0  &   66.8  &   60.9  &   46.5  & 0.51 $\pm$ 0.07  & 0.56 $\pm$ 0.08  & 0.73 $\pm$ 0.10  \\
 345.67  &    Fe\,{\sc{x}}  &   20.6  &   14.7  &   13.4  &   15.7  & 1.40 $\pm$ 0.20  & 1.54 $\pm$ 0.22  & 1.31 $\pm$ 0.19  \\
 347.34  &    Si\,{\sc{x}}  &   24.0  &   27.6  &   25.6  &   22.2  & 0.87 $\pm$ 0.12  & 0.94 $\pm$ 0.13  & 1.08 $\pm$ 0.15  \\
 352.58  &   Fe\,{\sc{xi}}   &   30.3  &   19.9  &   20.0  &   20.4  & 1.52 $\pm$ 0.22  & 1.51 $\pm$ 0.21  & 1.49 $\pm$ 0.21  \\
 368.11  & Mg\,{\sc{ix}}\tablenotemark{b} &  285.8  &  262.4  &  252.4  &  172.1  & 1.09 $\pm$ 0.15  & 1.13 $\pm$ 0.16  & 1.66 $\pm$ 0.23  
\enddata
\tablenotetext{a}{Column 1 is the wavelengths which are measured from the EUNIS-07 
LW spectrum. Column 2 is the ion name. Column 3 (I$_{e07}$) is the EUNIS-07 line
intensity. Column 4 (I$_{CDS}^{SN}$) is the CDS line intensity with the standard calibration and the new long-term correction. Column 5 (I$_{CDS}^{GZ}$) is the CDS
line intensity with the \citet{del01} calibration and the new long-term correction. Column 6 (I$_{CDS}^{S}$) is the CDS line intensity with the standard calibration and standard long-term correction.  Columns 7-9 are the EUNIS-to-CDS line intensity ratios.}
\tablenotetext{b}{The listed line intensity for Mg\,{\sc{ix}} 368.1 \AA\ includes the emission 
from the blended line, Mg\,{\sc{vii}} 367.7 \AA.}
\end{deluxetable*}

\subsection{SOHO/CDS}
The CDS includes a Normal Incidence Spectrometer (NIS) that can be used to obtain
stigmatic EUV spectra within its 308$-$381 \AA\ (NIS 1) and 513$-$633 \AA\ (NIS 2)
wavebands along its 240$^{''}$ long slit. Several slit widths are available, 
the most commonly used being 4$^{''}$. The instrument can be operated in 
a sit-and-stare mode or a rastering mode. In the latter case, the CDS scans 
a region of the Sun from the West to East without compensation for solar rotation, 
thus the actual FOV of a rastered spectroheliogram is stretched out in 
the x-direction when the targeted region is located on the solar disk.
In this study, the CDS rastered image was observed from 17:00:28 to 19:01:00 UT,
consisting of 60 pointing positions. The FOV is 260$^{''}\times240^{''}$. 
The pointing difference between CDS NIS 1 and NIS 2 has been corrected 
when applying {\it mk\_cds\_map} in SSW IDL library.

\subsection{Coalignments}
We first built up EUNIS rastered slit-lobe images in selected spectral lines 
for LW and SW channels using all exposure frames. Figure~\ref{fgfov} shows the 
LW He\,{\sc{ii}} 304 \AA\ slit-lobe image and the SW composite image from three 
coronal lines, Fe\,{\sc{x}} 174.5 \AA, Fe\,{\sc{x}} 177.2 \AA\ and Fe\,{\sc{xi}} 
180.4 \AA. We then determined the pointing, roll angle, and actual spatial scale 
of EUNIS slit images by coaligning them with EIT images in 304 \AA\ and 195 \AA\
passbands using the method as described in \citet{wan10}. For the EUNIS LW images 
we obtained the actual pixel size of 0$^{''}$.926~pixel$^{-1}$ and the roll angle 
of 3$^{\circ}$.47 counterclockwise from the North. For the SW images we obtained 
the actual pixel size of 0$^{''}$.920~pixel$^{-1}$ and the roll angle 
of 3$^{\circ}$.64. The measurements show that the LW and SW channels have 
nearly the same pixel size and roll angle, and they were well co-pointed 
in the direction perpendicular to the slit with a small offset of 
18$^{''}$.5 in solar x-direction and an offset of 108$^{''}$.5 in solar y-direction.
Figure~\ref{fgfov} shows the coaligned EUNIS-07 SW and LW slit-lobe images, 
indicating that their FOVs are mostly overlaid in the slit part. Figure~\ref{fgefv}
illustrates their cospatial rastered slit images in Si\,{\sc{x}} 347 \AA\ line 
of LW channel and Fe\,{\sc{xi}} 188 \AA\ line of SW channel.

The pointing of EIS LW band was determined by coaligning the He\,{\sc{ii}} 256 \AA\
rastered image obtained during 18:02--18:54 UT with the average image of EIT 304 \AA\ 
at 18:01 and 18:09 UT. The pointing of EIS SW band was determined by coaligning 
the Fe\,{\sc{xii}} 195 \AA\ raster image with the average image of EIT 195 \AA\ 
at 17:55 and 18:22 UT, where the average was made after the two EIT
images were rotated to a common time. The measured pointings for both
bands are the same considering the existing offsets between them. Figure~\ref{fgcfv}
illustrates the accurate coalignments between EUNIS-07 and EIS in both bands.

The pointing of CDS NIS was determined by coaligning the He\,{\sc{ii}} 304 \AA\ 
(second order) rastered image observed during 17:00$-$19:00 UT with the 
EIT 304 \AA\ image at 18:01 UT. Figure~\ref{fgcdsmap} illustrates a good coalignment 
between EUNIS and CDS NIS as seen in He\,{\sc{ii}} 304 \AA\ and Mg\,{\sc{ix}} 368 \AA\
rastered images. Comparisons between the FOVs of coaligned EUNIS, EIS, and 
CDS NIS are shown in Figure~\ref{fgfov}. 

 \begin{figure*}
 \epsscale{1.0}
 \plotone{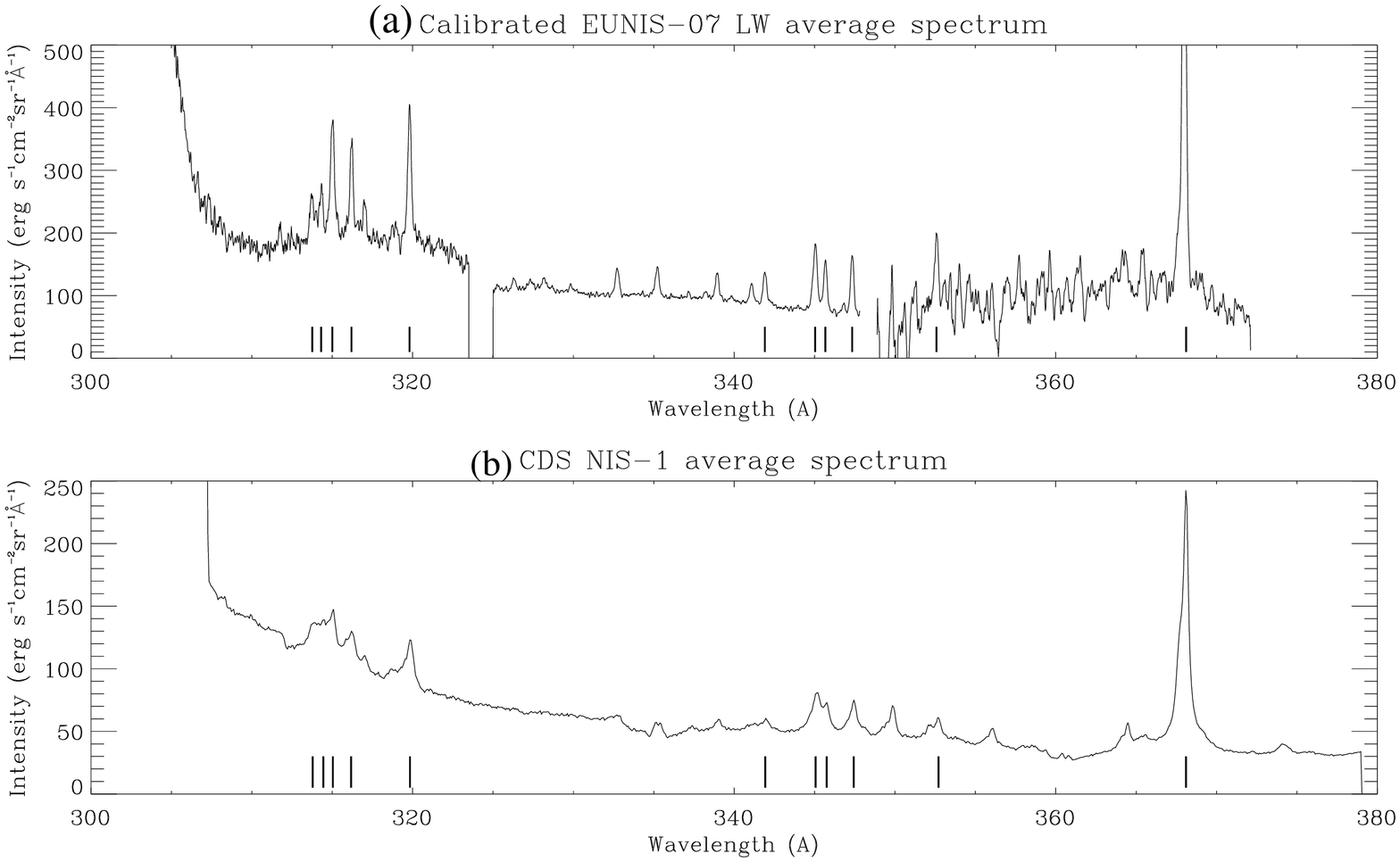}
 \caption{ \label{fgcdsspc}
  Averaged spectra for (a) EUNIS-07/LW channel and (b) CDS NIS-1 over their common 
FOV shown in Fig.~\ref{fgcdsmap}. The shown CDS spectrum was calibrated with the current
standard responsivities and the standard long-term corrections.
The bars in each panel mark the spectral lines which are used for cross-calibration between 
EUNIS and CDS, listed in 
Table~\ref{tabcds}.}
 \end{figure*}

 \begin{figure*}
 \epsscale{1.0}
 \plotone{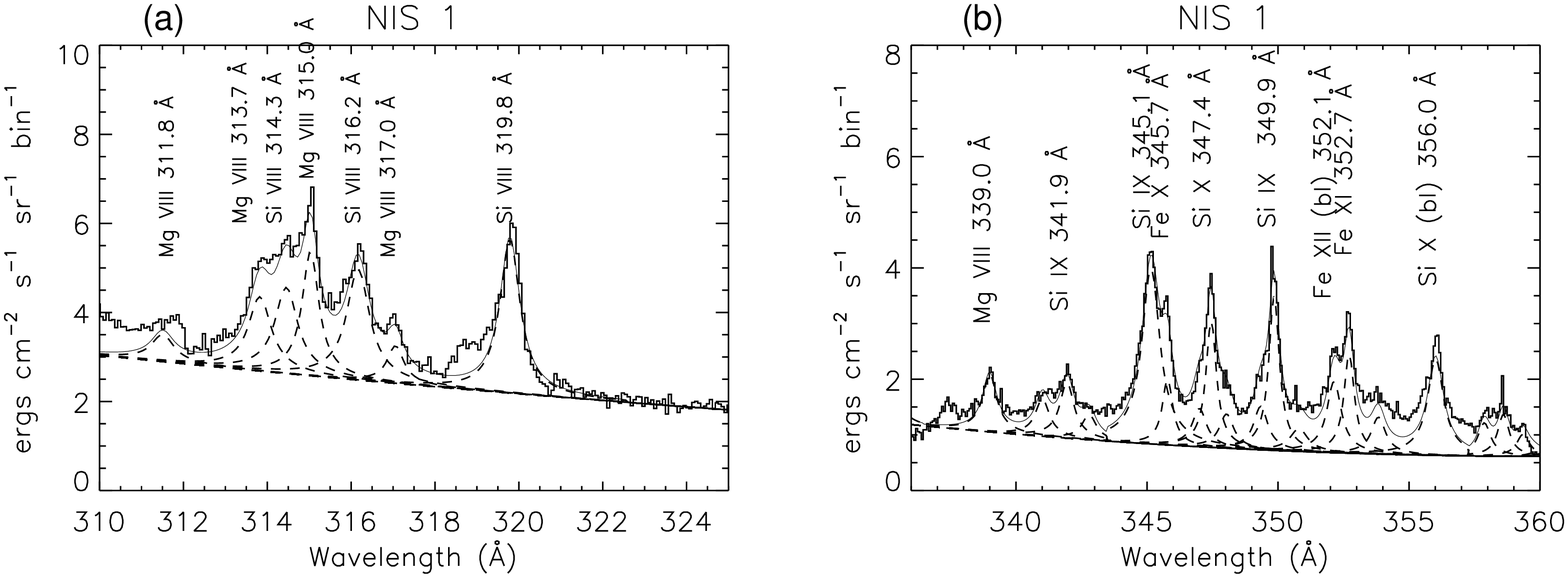}
 \caption{ \label{fgnisft}
  The broadened Gaussian fit to the CDS NIS-1 line profiles in the two selected windows of
(a) 310--325 \AA\ and (b) 336--360 \AA. The plotted NIS-1 spectrum was calibrated with the 
\citet{del01} responsivities and the new long-term corrections \citep{del10}. The spectrum
is shown in units of ergs~cm$^{-2}$~s$^{-1}$~sr$^{-1}$~bin$^{-1}$, where the ``bin" means
the CDS spectral pixel size ($\sim$0.0702 \AA). The obtained broadened Gaussian profiles and the 
polynomial background are shown in the dashed line, and their sum is in the solid line.}
 \end{figure*}

\begin{figure}
 \epsscale{1.0}
 \plotone{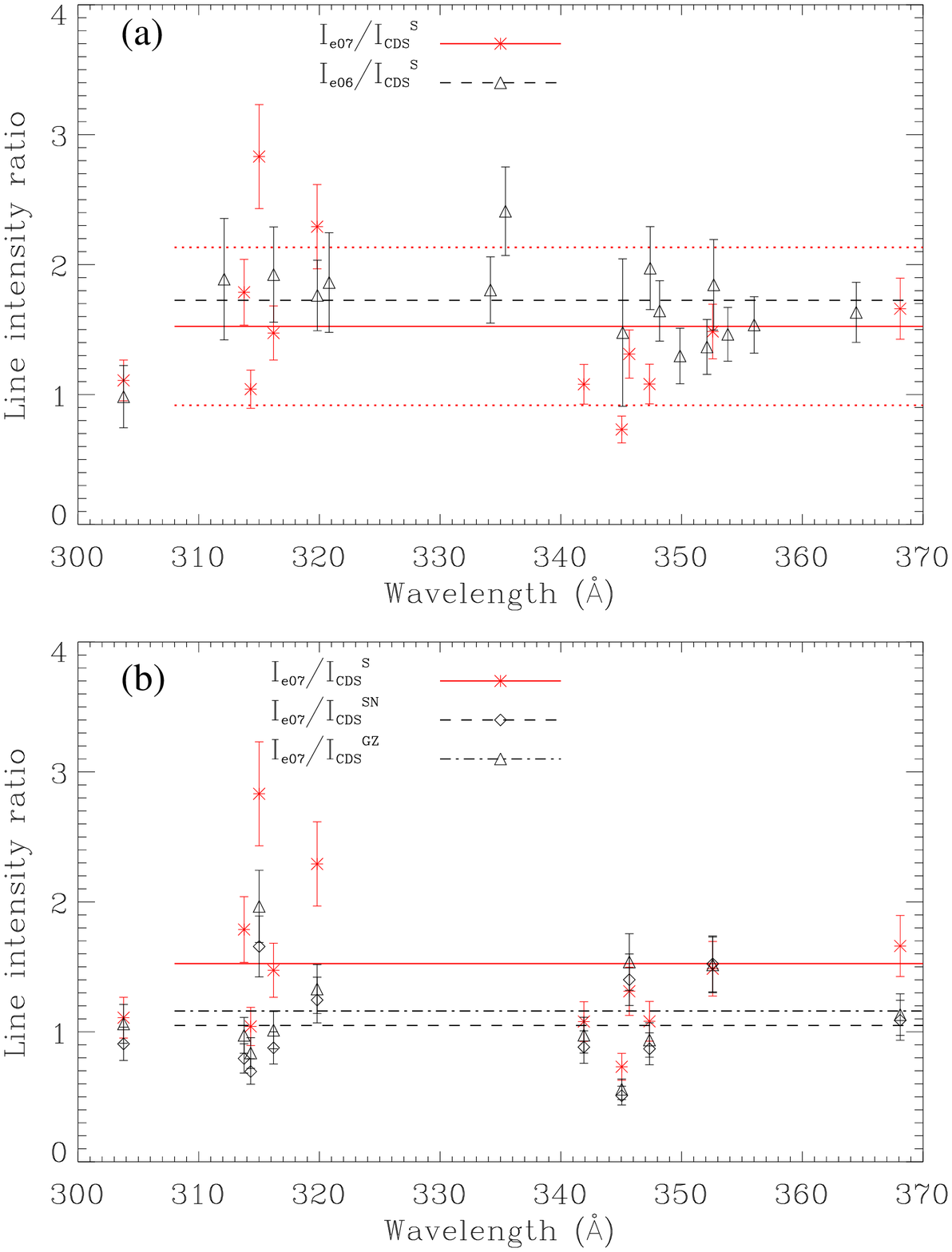}
 \caption{ \label{fgcdscal}
 (a) Plot of the EUNIS-to-CDS line intensity ratios. The ratios for EUNIS-07/CDS 
are denoted with {\it asterisks}, while those for EUNIS-06/CDS are with 
{\it triangles}. The data at 304 \AA\ correspond to the He\,{\sc{ii}} line ratio 
between EUNIS in first order and CDS NIS-2 in second order. The solid line 
represents the average for those lines in the wavelength range of 310$-$370 \AA\ 
for EUNIS-07, and the dotted lines are its standard deviation. The dashed line 
represents the average for the EUNIS-06/CDS line ratios. (b) Comparison between the EUNIS07-to-CDS
 line intensity ratios for the same CDS observation with different calibrations. 
The symbols of {\it asterisk} denote the ratios of the EUNIS intensity (I$_{e07}$) 
to the CDS intensity (I$_{CDS}^S$) obtained with the current standard CDS 
responsivities (version 4, 2002) and the standard long-term corrections. 
The symbols of {\it diamond} denote the ratios of the same EUNIS intensity to 
the CDS intensity (I$_{CDS}^{SN}$) obtained with the standard CDS responsivities 
and the new long-term corrections \citep{del10}. The symbols of {\it triangle} 
denote the ratios of the same EUNIS intensity to the CDS intensity (I$_{CDS}^{GZ}$)
obtained with the \citet{del01} scaled responsivities and the new long-term 
corrections. The solid line represents the average for ratios of I$_{e07}$/I$_{CDS}^S$,
the dashed line represents the average for ratios of I$_{e07}$/I$_{CDS}^{SN}$, and
the dot-dashed line represents the average for ratios of I$_{e07}$/I$_{CDS}^{GZ}$. }
 \end{figure}

\section{Absolute radiometric calibration of EUNIS-06 and EUNIS-07 LW channel}
Radiometric calibration of the EUNIS-06 LW channel was carried out in August 2006 
and that of the EUNIS-07 LW channel was in May 2008. They were performed at the RAL 
in the same facility and using the same EUV light source as was used for 
preflight calibrations of CDS \citep{lan02}. Recalibration
of the German PTB light source against the primary EUV radiation standard 
of BESSY-II ({\it English:} Berlin Electron 
Storage Ring Society for Synchrotron Radiation) in March 2007 showed 
that it had remained stable within its 7\% uncertainty. 

The end-to-end calibration of EUNIS-07 was made using the He\,{\sc{ii}} 304 \AA\ 
line at 88 individual locations, and using 11 distinct Ne features between 
300 and 370 \AA\ at 176 individual locations covering the instrument's entrance 
aperture. The aperture-averaged response at each wavelength was combined with 
the known source flux and with geometric factors, such as aperture area and 
pixel size, resulting in the absolute EUNIS responsivity within 
a total uncertainty of 10\% over its full LW bandpass. The spectral responsivity
measures the recorded data numbers per photon flux in ergs~cm$^{-2}$s$^{-1}$sr$^{-1}$. 
The measured absolute responsivities of EUNIS-07 and EUNIS-06 as a function 
of wavelength are shown in Figure~\ref{fglwc}. 
The shape of the response curve is mainly due to the 
multilayer coating on the spectrograph grating which is applied
to enhance its EUV efficiency. Error bars reflect the total range of values 
measured from the various entrance aperture positions. The fiber-optic coupling 
between the MCP and three active pixel sensors (APS) arrays in each
optical channel leads to differences in the overall responsivity of each APS array
relative to the other two. After correcting for these factors ($g_i$) which can be
measured in laboratory, the response curve ($R_{\lambda}$) in the range of 300--370 \AA\ of 
three APS arrays was obtained by a least-square parabolic fit to the measurements 
using
\begin{eqnarray}
 R_{\lambda}& = & g_i 10^{f(\lambda)}, \label{eqsen} \\ 
 f(\lambda)& = & a_0+a_1(\lambda-\lambda_0)+a_2(\lambda-\lambda_0)^2, \label{eqres} \\
g_i & = & \left\{ 
\begin{array}{ll}
1.000 & \quad \mbox{(300 $<\lambda<$324.8  \AA)} \label{eqfct}\\
3.107 & \quad \mbox{(324.8$<\lambda<$348.9 \AA)}\\
1.012 & \quad \mbox{(348.9$<\lambda<$370  \AA)}\\
\end{array}\right., 
\end{eqnarray}
where $f(\lambda)$ is the fitting function with $\lambda_0$=335 \AA\ and 
$R_{\lambda}$ is in units of REU (erg cm$^{-2}$sr$^{-1}$pixel$^{-1}$)$^{-1}$,
where the pixel unit is referred to as `spectral pixel' which is equal to 25 m\AA\ 
for first order spectrum. For EUNIS-07 we obtained the parameters, 
$a_0$=(8.0$\pm$4.8)$\times$10$^{-3}$, $a_1$=(4.3$\pm$0.1)$\times$10$^{-3}$, 
$a_2$=$-$(2.9$\pm$0.1)$\times$10$^{-4}$, while for EUNIS-06 obtained 
$a_0$=(1.8$\pm$0.1)$\times$10$^{-1}$, $a_1$=(4.2$\pm$0.4)$\times$10$^{-3}$, 
$a_2$=$-$(3.2$\pm$0.2)$\times$10$^{-4}$.
The shape of the response curve for EUNIS-07 has little change compared to 
that of EUNIS-06 (Figures~\ref{fglwc}(a) and~\ref{fglwc}(b)), but its absolute
responsivity has dropped by a factor of 1.45$\pm$0.06 (or decreased by $\sim$31\%) 
(Figure~\ref{fglwc}(c)). There was a slight shift in wavelength covered by
EUNIS-07 relative to EUNIS-06  due to a laboratory adjustment (see Fig.~\ref{fglwc}(b)). 
Validation of the laboratory calibration of the 
EUNIS-06 LW channel relevant to the shape of its response has been verified 
by checking its relative calibration using the density- and temperature-insensitive 
line ratio method \citep{wan10}. Therefore, no change in the shape of the response 
curve for EUNIS-07 supports the validation of its laboratory calibration.
In other words, because the insensitive ratio method confirmed the EUNIS-06
calibration, and the shape of the EUNIS-07 response curve remained constant 
relative to EUNIS-06, there is no need to repeat the insensitive ratio test.

\begin{deluxetable*}{llccccc}
 \tabletypesize{\scriptsize}
 \tablecaption{ Density- and temperature-insensitive line groups selected 
for EUNIS-07 SW calibration 
\label{tabswc}}
 \tablewidth{0pt}
 \tablehead{
 \colhead{Ion} &  \colhead{Wavelength} & \colhead{Theo. Ratio\tablenotemark{a}} & \colhead{Uncal. I$_{SW}$\tablenotemark{b}} & \colhead{Intensity\tablenotemark{c}} & \colhead{Abs. R$_{\lambda}$\tablenotemark{d}} &  \colhead{Rel. R$_{\lambda}$\tablenotemark{e}}\\
  \colhead{(1)} & \colhead{(2)} &  \colhead{(3)} & \colhead{(4)}  & \colhead{(5)} & \colhead{(6)} & \colhead{(7)}}
 \startdata
 Fe\,{\sc{x}}  & 345.74  & 1.00 $\pm$ 0.00  &  ...  &  22.90$ \pm$ 2.29  &    ...\\
        & 174.53  &  21.08 $\pm$ 3.04  &   1.21 $\pm$ 0.12  & 482.63 $\pm$ 84.70  &   2.51 $\pm$ 0.51  &   2.51 $\pm$ 0.51\\
        & 177.24  &  11.59 $\pm$ 1.54  &   0.81 $\pm$ 0.08  & 265.35 $\pm$ 44.13  &   3.05 $\pm$ 0.59  &   3.05 $\pm$ 0.59\\
        & 184.54  &   4.97 $\pm$ 0.19  &   1.57 $\pm$ 0.16  & 113.79 $\pm$ 12.18  &  13.75 $\pm$ 2.01  &   4.23 $\pm$ 0.62\\
\\
Fe\,{\sc{xi}} & 352.66  & 1.00 $\pm$ 0.00  &  ...  &  31.33 $\pm$ 3.13  &    ...\\
        & 180.41  &  11.44 $\pm$ 1.24  &   1.22 $\pm$  0.12  & 358.42 $\pm$ 52.86  &   3.40 $\pm$ 0.61  &   3.40 $\pm$ 0.61\\
        & 188.23\tablenotemark{f}  &   7.87 $\pm$ 0.24  &   3.29 $\pm$  0.33  & 246.57 $\pm$ 25.78  &  13.33 $\pm$ 1.93  &   4.10$\pm$  0.59\\
\\
Fe\,{\sc{xii}} & 364.47  & 1.00 $\pm$ 0.00  &  ...  &  21.04 $\pm$ 2.10  &    ...\\
        & 192.39  &   1.94 $\pm$  0.06  &   0.40 $\pm$  0.04  &  40.83 $\pm$  4.27  &   9.80 $\pm$  1.42  &   3.01 $\pm$  0.44\\
        & 193.51  &   4.06 $\pm$  0.15  &   0.93 $\pm$  0.09  &  85.44 $\pm$  9.11  &  10.84 $\pm$  1.58  &   3.33 $\pm$  0.49
 \enddata
\tablenotetext{a}{Theoretical line intensity ratios (relative to the LW line) 
calculated with CHIANTI package (ver.6).}  
\tablenotetext{b}{Uncalibrated SW line intensities in units of REU \AA\ s$^{-1}$.}
\tablenotetext{c}{The measured line intensity for LW and the derived line intensities 
for SW from the theoretical ratios in units of erg s$^{-1}$cm$^{-2}$sr$^{-1}$. }
\tablenotetext{d}{The absolute responsivity for SW in units of 10$^{-3}$ REU 
(erg cm$^{-2}$sr$^{-1}$\AA$^{-1}$)$^{-1}$ (Column 4/Column 5). }
\tablenotetext{e}{The relative responsivity with correction of APS relative 
sensitivity factors for the detectors, which has the same units as the absolute 
responsivity (see Eq.~(\ref{eqcsw})).}
\tablenotetext{f}{For Fe\,{\sc{xi}} 188.23, we have summed the intensities 
of the two Fe\,{\sc{xi}} lines at 188.23 and 188.30 \AA.}
\end{deluxetable*}

\section{Calibration update of CDS NIS with EUNIS-07}
The CDS NIS instrument was calibrated end-to-end at RAL, and various underflight
calibrations were implemented later on, forming the version 4
of the NIS `standard' calibration (implemented on May 21, 2002).
The history of the NIS calibration was described in detail by \citet{del10}.
\citet{del01} used the line-ratio technique to obtain an independent set
of responsivities which were consistent, to within a relative 30\%, with the `standard'
values (see the 2002 book on the SOHO calibration). The MCP in NIS detector
is known to have a drop in sensitivity over time due to exposure to solar radiation.
This results in a depression at the core of
the lines when one of the 2$^{''}$ or 4$^{''}$ slits is used (the so-called {\it burn-in}
of the lines). This effect in the NIS spectra has been monitored since launch with
the NIMCP study and is routinely corrected.
The use of the wide 90$^{''}$ slit was thought to be the main cause of the responsivity
loss over long time-scales \citep{tho06}. An empirical correction was implemented
within the standard software, and is still applied by default. However, \citet{del10} 
used the same NIMCP study to actually show that the burn-in due to the use
of the 90$^{''}$ slit is a secondary effect, while the responsivity in both NIS channels
has been steadily dropping. \citet{del10} provided a wavelength-dependent
long-term correction for this drop over the period 1996-2010.
The \citet{del10} long-term
corrections, combined with the \citet{del01} responsivities provided good
agreement between the irradiances of most CDS lines with those measured with the prototype
of the Solar Dynamics Observatory (SDO) EVE instrument, flown on a rocket on April 14, 2008.

The EUNIS LW channel has a wavelength range overlapping with that of CDS/NIS 1, 
so can provide a direct calibration update for it. The lab-calibrated SERTS-97 
was successfully used to improve the response curve for the CDS NIS 1 waveband 
based on their coordinated, cospatial spectra \citep{tho02}. Based on observations 
of active region spectra, EUNIS-06 provided a new calibration update for 
CDS NIS \citep{wan10}, which showed an overall decrease in NIS 1 responsivity 
by a factor about 1.7 compared to that of the previously implemented 
calibration (version 4, 2002) with the NIS ``standard" long-term corrections, 
while the responsivity in NIS 2 second order at 303$-$304 \AA\ with the NIS 
``standard" long-term corrections remained nearly constant (Fig.~\ref{fgcdscal}(a)).
It needs to be pointed out that the measured CDS He\,{\sc{II}} 304 intensity 
in \citet{wan10} was underestimated by a factor of 2 due to a bug found in 
the calibration routine, {\it nis\_calib}, which was fixed as of 2011 May. 
In the following we present the CDS underflight calibration update with 
EUNIS-07 measurements.

\begin{figure*}
 \epsscale{1.0 }
 \plotone{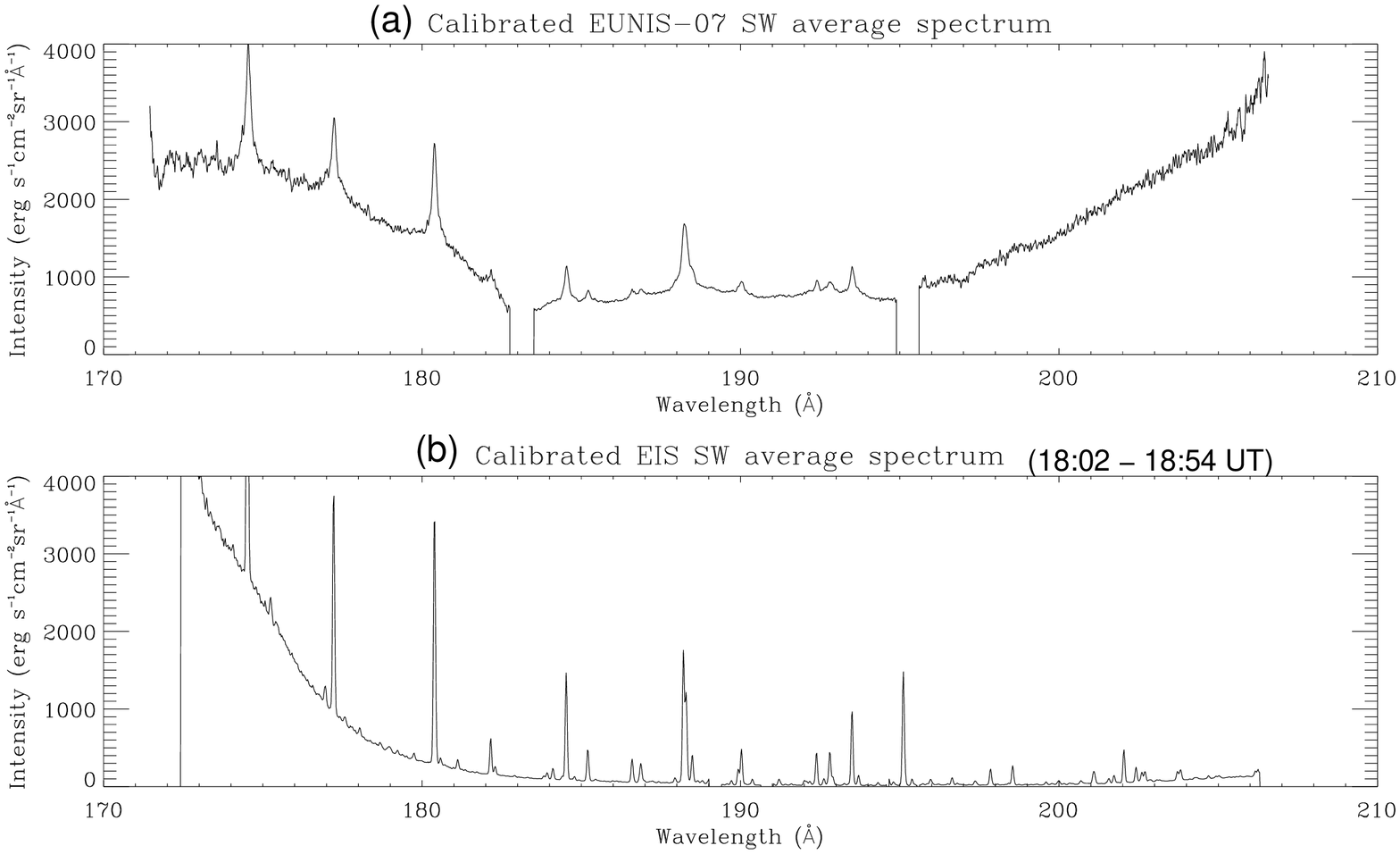}
 \caption{ \label{fgsws}
(a) Calibrated EUNIS-07 SW channel spectrum and (b) calibrated EIS SW channel 
spectrum, spatially averaged over a common FOV shown in Figures~\ref{fgcfv} (c) and (d).}
 \end{figure*}

Figure~\ref{fgcdsmap} shows the common FOVs between EUNIS-07 LW and CDS NIS as 
illustrated by the rastered images in He\,{\sc{ii}} 304 \AA\ and Mg\,{\sc{ix}} 368 \AA, 
indicating that a quiet region was observed. Figure~\ref{fgcdsspc} shows their 
cospatial average spectra, where 202 exposure frames were averaged for EUNIS-07.
Note that the two regions around the 335 and 360 \AA\ lines of CDS NIS 1 spectrum are
suffering from depression due to wide-slit burn-in effects from the Fe\,{\sc{xvi}} 
lines, where the long-term correction is more uncertain \citep{del10}. 
Both the calibrated EUNIS-07 LW and CDS NIS spectra (as well as the calibrated
EUNIS-07 SW and EIS spectra shown in Figure~\ref{fgsws} and Figure~\ref{fglws}) exhibit 
a significant background particularly in the low-responsivity wavelength range. This
mainly resulted from a combined effect of the instrumental noise and scattered light with
the special shape of their response curves. For the EUNIS spectra, we first divided the
full spectral window into several small sections, and then fitted spectral lines by 
a multiple Gaussian function with a linear background using the standard line 
fitting routine ({\it xcfit.pro}) in SSW. In this way the slowly, smoothly varying 
(with wavelength) ``background'' level only slightly affects the integrated intensities 
of the fitted line profiles. Measurements of the line intensities from EUNIS-07 SW and
EIS SW/LW spectra in Sections~\ref{sctswc} and~\ref{scteis} were made using the 
same technique. The CDS line intensities were obtained by fitting  multiple 
broadened Gaussian functions to the full spectrum. The broadened Gaussian function 
is a combination of a Gaussian term plus a term describing the wings \citep{thomp99}.
We chose the broadening with no asymmetry and kept the relative amplitude of 
the wings equal to a value of 0.8, but allowed the width of the Gaussian term for 
the line to vary (within 0.29 and 0.4~\AA\ in FWHM), as well as its
centroid and peak values. All lines were fitted at once, with the background fitted 
with a polynomial curve. Figure~\ref{fgnisft} shows the fitted CDS line profiles in
two selected wavelength ranges. Even with its moderate spectral resolution,
CDS can resolve the lines that we used for calibration.  The uncertainty caused by 
the moderate spectral resolution of CDS is relatively small compared to that 
in the background when assuming the common background fit for all spectral lines.

Table~\ref{tabcds} lists 12 strong emission lines observed by both EUNIS LW and CDS 
(NIS 1 in first order and NIS 2 in second order). To examine the current CDS 
responsivities and long-term corrections, we measured the line intensities with 
three kinds of calibration as used in \citet{del10}. The fourth column ($I_{CDS}^{SN}$)
shows the intensities obtained with the current standard CDS responsivities 
and the new long-term corrections derived by \citet{del10}, 
which has been implemented with the keyword {\it alt\_slit6} in the SSW routine
{\it vds\_calib}. The fifth column ($I_{CDS}^{GZ}$) shows the intensities obtained 
with the \citet{del01} scaled responsivities (using {\it nis\_resp\_gdz\_2010} in SSW)
and the new long-term corrections. Note that differences between the standard CDS
responsivities and the \citet{del01} scaled ones are small, overall of the order of 
10\% or so. The sixth column ($I_{CDS}^{S}$) shows the intensities obtained 
with the current standard CDS responsivities (version 4, 2002) and the standard 
long-term corrections (using {\it vds\_calib} with the keyword {\it slit6} [default] ). 
Considering long-term corrections have the relative uncertainty of 20\%, combined with
the uncertainties in the background and the responsivities 
\citep[20\% to 30\% see][]{del01}, it means a typical uncertainty of 30\%--40\% in
measurements of CDS line intensities. Ratios of the EUNIS-07 to various CDS line
intensities are listed in columns 7$-$9 and shown in Figure~\ref{fgcdscal}, 
where errors of the ratios were calculated by assuming 10\% uncertainty (as a lower limit)
in measurements of both EUNIS and CDS line intensities. 

 \begin{figure*}
 \epsscale{1.0}
 \plotone{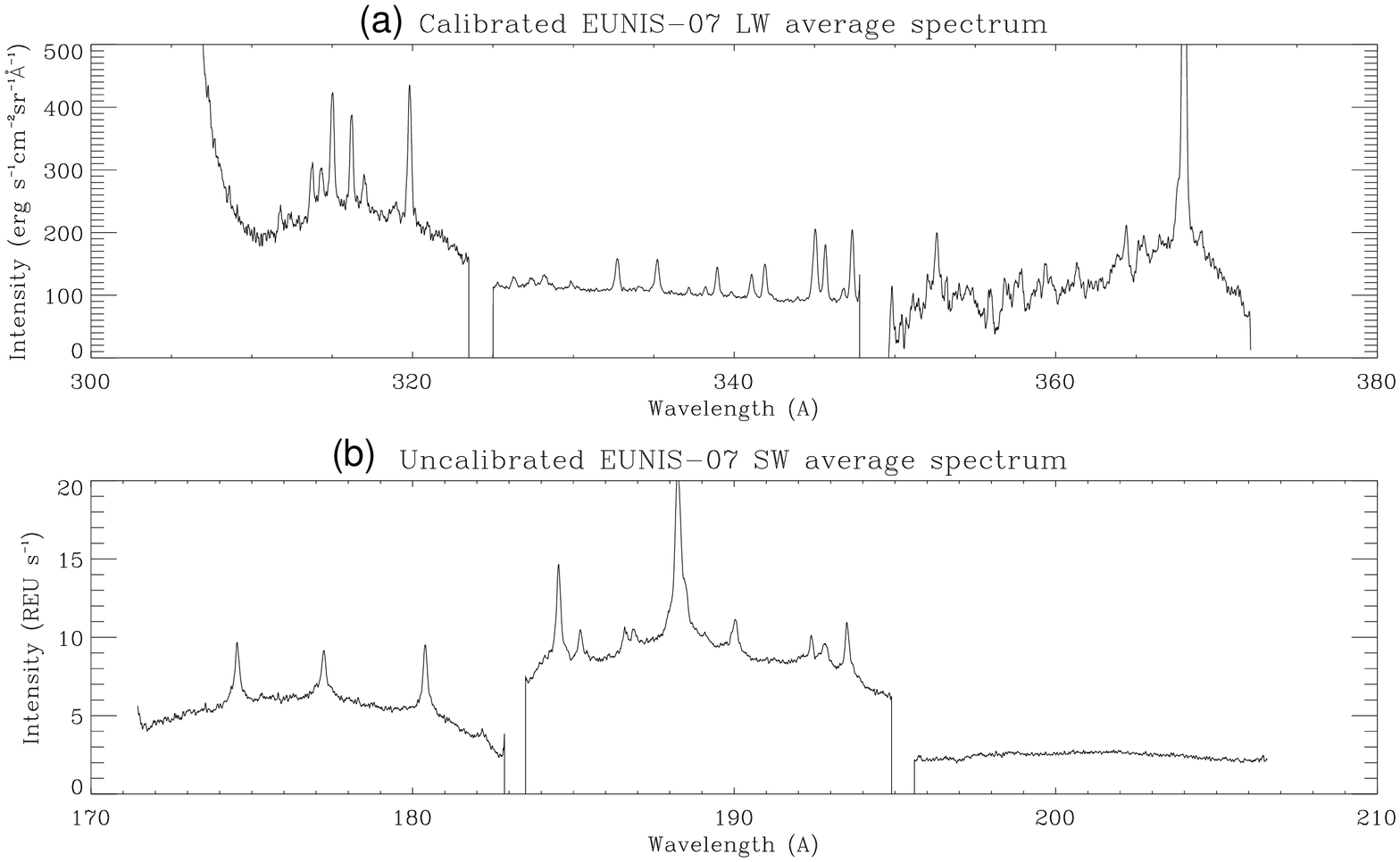}
 \caption{ \label{fgesp}
(a) Calibrated EUNIS-07/LW channel spectrum and (b) uncalibrated EUNIS-07/SW 
channel spectrum, averaged over pixels in a common FOV for 202 exposures. }
 \end{figure*}

 \begin{figure}
 \epsscale{1.0}
 \plotone{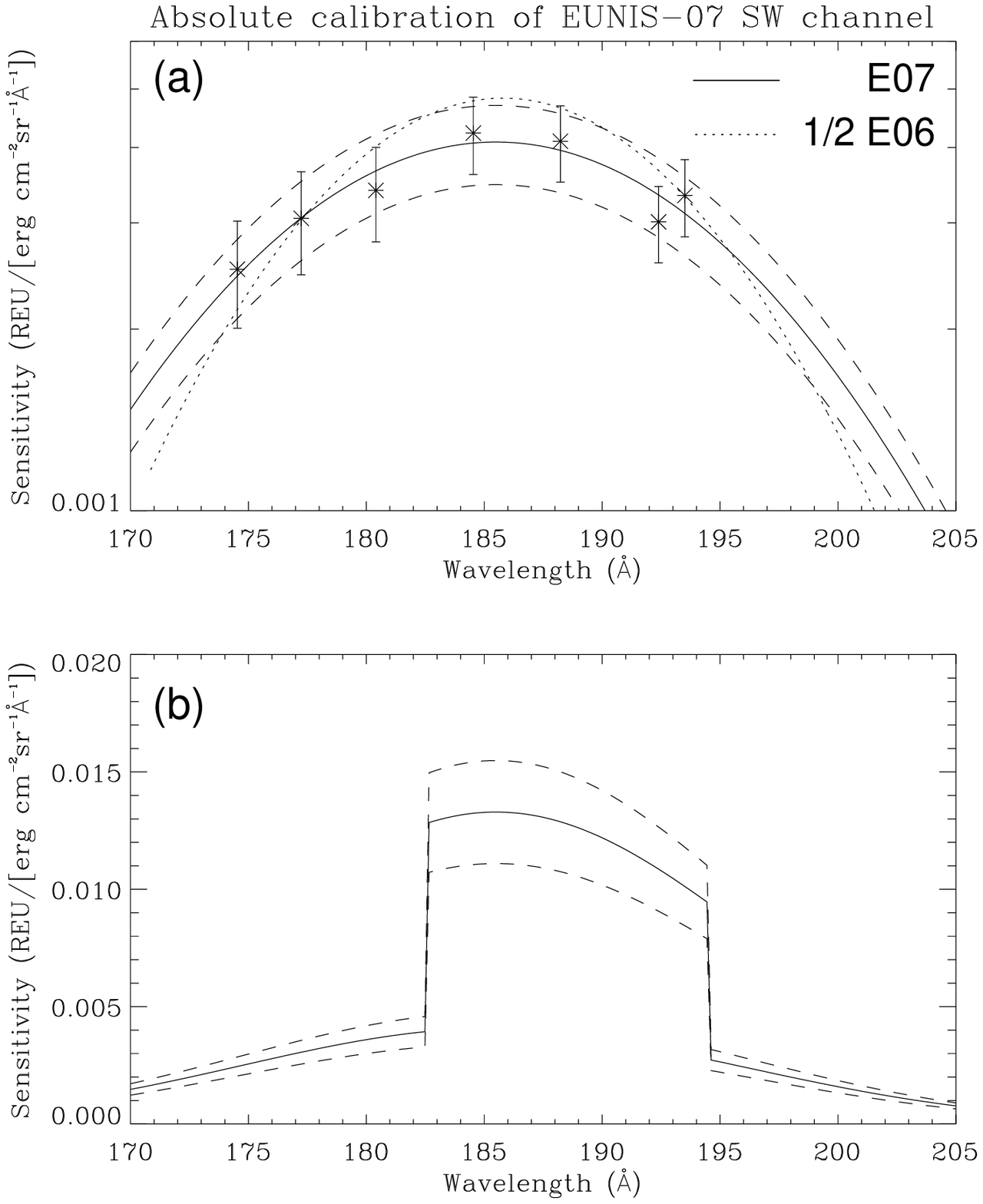}
 \caption{ \label{fgucl}
 (a) Measured instrument responsivity for EUNIS-07 SW channel using density- and
 temperature-insensitive line intensity ratios. The thick solid curve is 
a least-squares parabolic fit to the data points. The dashed lines indicate 
the 15\% uncertainty. The dotted line represents the measured instrument responsivity 
for EUNIS-06 SW which has been reduced by a factor of 2 for comparison of 
their shapes. (b) The same calibration curves as in (a) but on a linear scale 
and without correction of relative sensitivity factors for the detectors.  }
 \end{figure}

The differences are significant between the EUNIS-07 line intensities and the CDS 
ones applied with standard responsivities and standard long-term corrections, 
especially for those lines in the wavelength range of 310$-$320 \AA. The NIS 1
responsivity (with standard long-term corrections) overall decreased by a factor 
of 1.5$\pm$0.6, which is consistent with that derived by EUNIS-06.  
While the EUNIS-07 line intensities are well consistent  with 
the CDS ones applied with new long-term corrections in many lines (with differences
less than 15\%). Their average ratio is 1.05$\pm$0.36 in the case of
the standard responsivities and it is 1.16$\pm$0.39 in the case of the 
\citet{del01} scaled ones. Therefore, the results of EUNIS-06 and EUNIS-07 flights
confirmed the new long-term corrections for NIS 1 response obtained by \citet{del10}, 
and strongly suggest the application of keyword, {\it alt\_slit6}, 
in using {\it vds\_calib} or {\it nis\_calib} to calibrate CDS data.
In addition, our measurements show that the differences are small (on the order 
of 10\%) between He\,{\sc{ii}} 304 \AA\ intensities obtained by EUNIS-07 in first order 
and by CDS NIS-2 in second order applied with the standard or new long-term 
corrections, indicating that the second-order standard (version 4) responsivity 
of NIS 2 updated by SERTS-97 \citep{tho02} and EGS \citep{woo98} remains to work well 
(with a decrease in uncertainty) when the standard long-term corrections are applied.    
Finally, we notice that even after the new long-term corrections, the CDS
intensities for some lines showed large ($\gtrsim$30\%) difference from the EUNIS 
ones. This could be due to the uncertainties involved 
in difficulty of background (scattered light) estimate for the CDS spectrum.

\begin{deluxetable*}{llcccccc}
 \tabletypesize{\scriptsize}
 \tablecaption{ Absolutely calibrated quiet-Sun region line lists for EUNIS-07 SW 
channel and EIS/SW band, and the derived instrument responsivity for EIS SW band 
from the measured EUNIS absolute radiometric calibration. 
\label{tabeis}}
 \tablewidth{0pt}
 \tablehead{
 \colhead{Ion} &  \colhead{Wavelength} & \colhead{I (Uncal.E07)\tablenotemark{a}} & \colhead{I (Cal.E07)\tablenotemark{b}} & 
        \colhead{I (Uncal.EIS)\tablenotemark{c}} & \colhead{I (Cal.EIS)\tablenotemark{d}} &  \colhead{I$_{E07}$/I$_{EIS}$\tablenotemark{e}}  
       & \colhead{EIS R$_{\lambda}$\tablenotemark{f}}\\
 \colhead{(1)}  & \colhead{(2)} & \colhead{(3)} & \colhead{(4)} & \colhead{(5)} & \colhead{(6)} & \colhead{(7)} & \colhead{(8)} }
 \startdata
Fe\,{\sc{x}}  & 174.54  &1.281 $\pm$ 0.128  & 522.37 $\pm$ 52.24  & 0.799 $\pm$ 0.080  & 393.03 $\pm$ 39.30  &1.329 $\pm$ 0.188  &  1.53e-03 $\pm$ 2.16e-04\\
Fe\,{\sc{x}}  & 177.24  &0.799 $\pm$ 0.080  & 261.56 $\pm$ 26.16  & 1.313 $\pm$ 0.131  & 212.94 $\pm$ 21.29  &1.228 $\pm$ 0.174  &  5.02e-03 $\pm$ 7.10e-04\\
Fe\,{\sc{xi}} & 180.39  &1.261 $\pm$ 0.126  & 344.89 $\pm$ 34.49  & 5.530 $\pm$ 0.553  & 257.12 $\pm$ 25.71  &1.341 $\pm$ 0.190  &  1.60e-02 $\pm$ 2.27e-03\\
Fe\,{\sc{x}} & 184.54  &1.617 $\pm$ 0.162  & 122.16 $\pm$ 12.22  & 8.524 $\pm$ 0.852  & 102.33 $\pm$ 10.23  &1.194 $\pm$ 0.169  &  6.98e-02 $\pm$ 9.87e-03\\
Fe\,{\sc{viii}} & 185.22  &0.484 $\pm$ 0.048  &  36.46 $\pm$ 3.65  & 3.035 $\pm$ 0.303  &  30.48 $\pm$ 3.05  &1.196 $\pm$ 0.169  &  8.32e-02 $\pm$ 1.18e-02\\
Fe\,{\sc{viii}}  & 186.60  &0.335 $\pm$ 0.034  &  25.35 $\pm$ 2.54  & 3.229 $\pm$ 0.323  &  23.41 $\pm$ 2.34  &1.083 $\pm$ 0.153  &  1.27e-01 $\pm$ 1.80e-02\\
Fe\,{\sc{xii}} & 186.88  &0.318 $\pm$ 0.032  &  24.13 $\pm$ 2.41  & 3.204 $\pm$ 0.320  &  21.93 $\pm$ 2.19  &1.100 $\pm$ 0.156  &  1.33e-01 $\pm$ 1.88e-02\\
Fe\,{\sc{xi}} & 188.23\tablenotemark{g}  &3.515 $\pm$ 0.351  & 272.88 $\pm$ 27.29  &39.602 $\pm$ 3.960  & 209.95 $\pm$ 21.00  &1.300 $\pm$ 0.184  &  1.45e-01 $\pm$ 2.05e-02\\
Fe\,{\sc{x}}  & 190.04  &0.676 $\pm$ 0.068  &  55.53 $\pm$ 5.55  &12.383 $\pm$ 1.238  &  49.60 $\pm$ 4.96  &1.120 $\pm$ 0.158  &  2.23e-01 $\pm$ 3.15e-02\\
Fe\,{\sc{xii}} & 192.40  &0.444 $\pm$ 0.044  &  40.93 $\pm$ 4.09  &10.596 $\pm$ 1.060  &  32.49 $\pm$  3.25  &1.260 $\pm$ 0.178  &  2.59e-01 $\pm$ 3.66e-02\\
Fe\,{\sc{xii}} & 193.51  &0.959 $\pm$ 0.096  &  94.80 $\pm$ 9.48  &26.673 $\pm$ 2.667  &  74.79 $\pm$  7.48  &1.268 $\pm$ 0.179  &  2.81e-01 $\pm$ 3.98e-02
 \enddata
\tablenotetext{a}{Uncalibrated EUNIS-07 SW line intensities in units of 
REU \AA\ s$^{-1}$.}
\tablenotetext{b}{Calibrated EUNIS-07 SW line intensities in units of 
erg s$^{-1}$cm$^{-2}$sr$^{-1}$.}
\tablenotetext{c}{Uncalibrated EIS SW line intensities in units of 
DN spec\_pix s$^{-1}$.}
\tablenotetext{d}{Calibrated EIS SW line intensities in units of 
erg s$^{-1}$cm$^{-2}$sr$^{-1}$.}
\tablenotetext{e}{The ratio of calibrated EUNIS-07 SW line intensities to 
the EIS SW line intensities.}
\tablenotetext{f}{The derived responsivity for EIS SW in units of 
DN spec\_pix (erg cm$^{-2}$sr$^{-1}$)$^{-1}$ (Column 5/Column 4).}
\tablenotetext{g}{For Fe\,{\sc{xi}} 188.23, we have summed the intensities 
of the two Fe\,{\sc{xi}} lines at 188.23 and 188.30 \AA.}
\end{deluxetable*}

\section{Absolute radiometric calibration of EUNIS-07 SW channel}
\label{sctswc}
The insensitive line ratio method was proposed by \citet{neu83} as a means of 
monitoring relative calibration variations of inflight EUV spectrometers and was 
used by \citet{tho94} and \citet{bro96} to adjust the laboratory calibration curve 
for SERTS-89, SERTS-91 and SERTS-93. \citet{bro98a, bro98b} derived the SERTS-95
relative radiometric calibration for both the first-order and second-order wave 
bands with this technique. In the following we used the lab-calibrated EUNIS-07 LW
channel to calibrate its SW channel by means of density- and temperature-insensitive 
line intensity ratios because the responsivity of its SW channel could not be 
directly measured at RAL.

Figure~\ref{fgefv} shows the common FOV between LW and SW channels of EUNIS-07 
as illustrated by the rastered images in coronal lines,  Si\,{\sc{x}} 347.4 \AA\ 
from LW and Fe\,{\sc{xi}} 188 (188.2$+$188.3) \AA\ from SW. 
Figure~\ref{fgesp} shows the spectra 
averaged over the same spatial area for 202 exposure frames observed from 
18:02:14 to 18:06:34 UT. Line intensities of the EUNIS LW spectrum were measured 
with the Gaussian fit, while those of the SW spectrum with the broadened Gaussian fit.
Table~\ref{tabswc} lists three groups of emission lines for Fe\,{\sc{x}},
Fe\,{\sc{xi}}, and Fe\,{\sc{xii}}. Column (3) gives theoretical line intensities 
relative to the one in the wavelength range for LW, which were calculated with 
the CHIANTI ver.~6 package \citep{der97, der09}. The line ratios exhibit slight
variations with electron density ($N_e$), so the mean value over a range of 
$8.5\leq{\rm log}_{10}N_e\leq{10.5}$ was taken, with the uncertainty corresponding
to half of the difference between the maximum and minimum values. In each case, 
the intensity ratios were calculated at the temperature of maximum ion abundance 
(i.e., the lines' formation temperature), using the ionization equilibrium data 
({\it chianti.ioneq}). Column (4) gives the uncalibrated line intensity of 
SW spectral lines. In Column (5) the absolute intensity for LW lines was 
directly measured from the average spectrum, while the absolute intensity of SW lines 
was derived from the measured LW line intensity and CHIANTI theoretical insensitive 
line ratio.  Column (6) gives the absolute responsivities, which are the ratio of 
the measured uncalibrated intensity (Column (4)) to the derived calibrated 
intensity (Column (5)) for SW lines. Column (7) gives the relative responsivities 
which were derived from the absolute ones by correcting for the relative sensitivity
factors ($g_i$). Figure~\ref{fgucl}(a) shows the relative responsivities and 
a least-squares parabolic fit on a logarithmic scale. The radiometric calibration 
response curve ($R_{\lambda}$) in the range of 170$-$205 \AA\ is shown in 
Figure~\ref{fgucl}(b), obtained by Equations~(\ref{eqsen}) and~(\ref{eqres}),
where $f(\lambda)$ is the fitting function with $\lambda_0$=187.5 \AA, 
$a_0$=$-$2.40$\pm$0.04, $a_1$=$-$(7.4$\pm$5.9)$\times$10$^{-3}$, 
$a_2$=$-$(1.8$\pm$0.8)$\times$10$^{-3}$, and the relative sensitivity factors 
for the three APS arrays were determined in laboratory as
\begin{equation}
g_i = \left\{ \label{eqcsw}
\begin{array}{ll}
1.000 & \quad \mbox{(170 $<\lambda<$182.5  \AA)}\\
3.254 & \quad \mbox{(182.5$<\lambda<$194.5 \AA)}\\
0.950 & \quad \mbox{(194.5$<\lambda<$205  \AA)}\\
\end{array}\right..
\end{equation}
The $R_{\lambda}$ has units of REU (erg cm$^{-2}$sr$^{-1}$\AA$^{-1}$)$^{-1}$.
The shape of the SW response curve of EUNIS-07 is consistent with that of EUNIS-06 
within the 15\% uncertainty, but its responsivity has decreased by a factor of 2. 
Note that because the responsivity had fallen significantly and the observed region was 
very quiet, no emission lines were observed by the right ASP detector in SW channel 
which covers the wavelength range from 196 to 207 \AA, including three Fe\,{\sc{xiii}}
(200.0, 203.7, 202.0 \AA) lines that had been used for EUNIS-06 SW radiometric 
calibration. Therefore, the responsivities in the wavelength range between 194 and 205 
\AA\ are not directly constrained by observations.

\begin{deluxetable*}{llcccc}
 \tabletypesize{\scriptsize}
 \tablecaption{ Density- and temperature-insensitive line groups selected 
for EIS calibration.
\label{tabrat}}
 \tablewidth{0pt}
 \tablehead{\colhead{Ion} & \colhead{Wavelength} & \colhead{Theo. Ratio\tablenotemark{a}} & \colhead{I$_{E07}$\tablenotemark{b}} & \colhead{I$_{EIS}$\tablenotemark{c}} & \colhead{I$_{E07}$/I$_{EIS}$\tablenotemark{d}} \\
 \colhead{(1)} & \colhead{(2)} &  \colhead{(3)} & \colhead{(4)}  & \colhead{(5)} & \colhead{(6)} }
\startdata
Fe\,{\sc{x}}  & 345.74  & 1.00$\pm$0.00  &  23.35$\pm$2.34  &  ...  &  ...\\
        & 174.53  &  21.07$\pm$  3.04  & 491.98$\pm$ 86.37  & 389.96$\pm$ 39.00  &   1.26$\pm$  0.25\\
        & 177.24  &  11.58$\pm$  1.54  & 270.39$\pm$ 44.99  & 209.59$\pm$ 20.96  &   1.29$\pm$  0.25\\
        & 184.54  &   4.97$\pm$  0.19  & 116.05$\pm$ 12.42  & 101.55$\pm$ 10.16  &   1.14$\pm$  0.17\\
Fe\,{\sc{xi}}  & 352.66  & 1.00$\pm$0.00  &  35.25$\pm$3.77  &  ...  &  ...\\
        & 188.23\tablenotemark{e}  &   7.87$\pm$  0.24  & 277.42$\pm$ 30.85  & 212.92$\pm$ 21.29  &   1.30$\pm$  0.19\\
        & 192.83\tablenotemark{f}  &   1.20$\pm$  0.06  &  42.30$\pm$ 4.99  &  34.99$\pm$  3.50  &   1.21$\pm$  0.19\\
Fe\,{\sc{xi}}  & 341.11  & 1.00$\pm$0.00  &  10.73$\pm$1.08  &  ...  &  ...\\
        & 188.23\tablenotemark{e}  &  28.33$\pm$  4.67  & 303.98$\pm$ 58.64  & 212.92$\pm$ 21.29  &   1.43$\pm$  0.31\\
        & 192.83\tablenotemark{f}  &   4.31$\pm$  0.77  &  46.25$\pm$ 9.47  &  34.99$\pm$  3.50  &   1.32$\pm$  0.30\\
Fe\,{\sc{xii}}  & 352.11  & 1.00$\pm$0.00  &  13.10$\pm$3.37  &  ...  &  ...\\
        & 192.39  &   3.22$\pm$  0.14  &  42.17$\pm$ 11.01 &  32.86$\pm$  3.29  &   1.28$\pm$  0.36\\
        & 193.51  &   6.75$\pm$  0.34  &  88.39$\pm$ 23.18  &  76.74$\pm$  7.67  &   1.15$\pm$  0.32\\
        & 195.12  &  10.63$\pm$  0.19  & 139.20$\pm$ 35.91  & 117.14$\pm$ 11.71  &   1.19$\pm$  0.33\\
Fe\,{\sc{xii}} & 364.47  & 1.00$\pm$0.00  &  21.07$\pm$2.69  &  ...  &  ...\\
        & 192.39  &   1.94$\pm$  0.06  &  40.88$\pm$ 5.37  &  32.86$\pm$  3.29  &   1.24$\pm$  0.21\\
        & 193.51  &   4.06$\pm$  0.15  &  85.54$\pm$ 11.37  &  76.74$\pm$  7.67  &   1.11$\pm$  0.19\\
        & 195.12  &   6.39$\pm$  0.21  & 134.64$\pm$ 17.75  & 117.14$\pm$ 11.71  &   1.15$\pm$  0.19\\
Fe\,{\sc{xii}}  & 338.26  & 1.00$\pm$0.00  &   3.28$\pm$0.33  &  ...  &  ...\\
        & 186.85\tablenotemark{g}  &   9.32$\pm$  1.94  &  30.52$\pm$7.06  &  24.08$\pm$  2.41  &   1.27$\pm$  0.32\\
Si\,{\sc{x}}  & 347.41  & 1.00$\pm$0.00  &  30.90$\pm$3.09  &  ...  &  ...\\
        & 261.04  &   0.79$\pm$  0.12  &  24.41$\pm$4.44   &  19.31$\pm$  1.93  &   1.26$\pm$  0.26\\
        & 277.28  &   0.48$\pm$  0.07  &  14.83$\pm$ 2.6  &  11.63$\pm$  1.16  &   1.28$\pm$  0.26\\
        & 272.01  &   0.59$\pm$  0.09  &  18.23$\pm$ 3.32  &  17.03$\pm$  1.70  &   1.07$\pm$  0.22
\enddata
\tablenotetext{a}{Theoretical line intensity ratios (relative to the EUNIS-07 LW line)
 calculated with CHIANTI package (ver.6).}
\tablenotetext{b}{For wavelengths between 330 and 370 \AA, the absolute intensity 
(in erg s$^{-1}$cm$^{-2}$sr$^{-1}$) was measured from the calibrated EUNIS-07 
LW spectrum, while for  wavelengths between 170 and 280 \AA, the absolute intensity 
was derived from the EUNIS-07 LW line intensity and CHIANTI ver.6.}
\tablenotetext{c}{The measured absolute intensities 
(in erg s$^{-1}$cm$^{-2}$sr$^{-1}$) for EIS SW and LW insensitive lines.}
\tablenotetext{d}{The EUNIS-07 to EIS line intensity ratios.}
\tablenotetext{e}{For Fe\,{\sc{xi}} 188.23, the intensities of two self-blended lines 
 at 188.23 and 188.30 \AA\ were summed.}
\tablenotetext{f}{For Fe\,{\sc{xi}} 192.83 \AA, the blending from a weak line O\,{\sc{v}} 192.8 \AA\
 could lead to the derived EUNIS-to-EIS intensity ratio underestimated by about 7\%. }
\tablenotetext{g}{For Fe\,{\sc{xii}} 186.85 \AA, the intensities of two self-blended lines
 at 186.85 and 186.89 \AA\ were summed.}
\end{deluxetable*}

 \begin{figure}
 \epsscale{1.0}
 \plotone{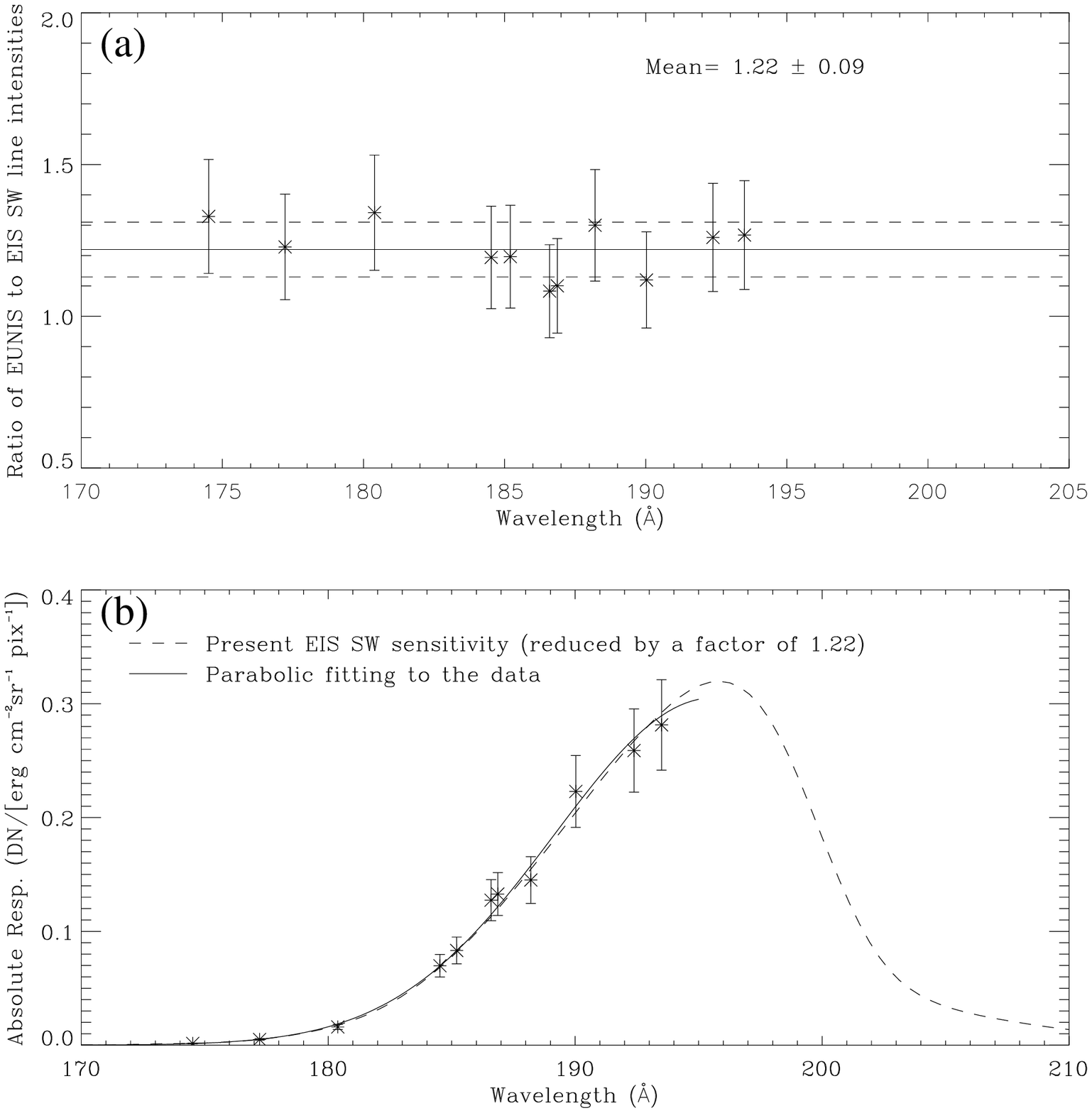}
 \caption{ \label{fgecl}
 (a) Plot of the EUNIS-07/SW to EIS/SW intensity ratios for the spectral lines listed 
in Table~\ref{tabeis} (column 7). The solid horizontal line represents the average ratio and 
the dashed lines their standard deviation. (b) Measured EIS/SW band responsivity using 
the absolute radiometric calibration of EUNIS-07. The thick solid curve is 
a least-squares parabolic fit to the data points (column 8). The dashed curve is the present
EIS/SW responsivity used in SSW, which has been reduced by a factor of 1.22 for comparison.}
 \end{figure}

\section{Calibration update of EIS with EUNIS-07}
\label{scteis}
We determined the EIS radiometric calibration using two methods. One is based on direct
transfer of the radiometric calibration for EUNIS-07 SW channel, whose wavelength 
range overlaps that of the EIS SW band. The other is based on the temperature- and 
density-insensitive line ratio technique, in which one member of each insensitive 
line pair was observed by EUNIS-07 LW channel, and the other by EIS LW or SW band. In the
following we applied the first method to the EIS raster observed after the EUNIS-07 
flight, and applied the second method to the EIS rasters observed both before and
after the EUNIS-07 flight.  

\subsection{Method 1: Calibration transfer from EUNIS-07 SW to EIS SW}
Figure~\ref{fgsws} shows the cospatial average spectra of EUNIS-07 SW and EIS SW bands
over a common FOV shown in Figures~\ref{fgcfv}(c) and~\ref{fgcfv}(d). The EUNIS SW
spectrum was calibrated with the response curve determined in Section~\ref{sctswc}. 
The EIS spectrum was calibrated using the standard routine, {\it eis\_prep} in SSW, 
which used the pre-launch radiometric calibration based on laboratory measurements 
\citep{see04, lan06, cul07, brw08}. The EIS average spectrum was fitted using 
an interactive multi-Gaussian (with a linear background) fitting routine, 
{\it spec\_gauss\_eis} in SSW. Table~\ref{tabeis} lists the measured calibrated 
and uncalibrated line intensities for 11 strong emission lines of Fe\,{\sc{viii}}, 
Fe\,{\sc{x}}, Fe\,{\sc{xi}}, and Fe\,{\sc{xii}}. A 10\% uncertainty was assumed
for the measured EUNIS and EIS line intensities because the statistical errors in 
the fitting were negligible, while the uncertainties due to absolute radiometric 
calibration may be up to 20\%. Column (7) gives the EUNIS-to-EIS
line intensity ratios (see Figure~\ref{fgecl}(a)). Column (8) gives the derived EIS
responsivities which are ratios of the uncalibrated EIS to calibrated EUNIS line 
intensities (see Figure~\ref{fgecl}(b)). Our measurements indicate that the EIS
responsivity decreased by a factor of 1.22$\pm$0.09 after a year inflight. The shape of 
the derived EIS response curve is well consistent with the one measured before its
launch. The EIS response curve 
(in units of DN (erg cm$^{-2}$sr$^{-1}$pixel$^{-1}$)$^{-1}$) 
derived with a least-squares parabolic fit on a logarithmic scale in the wavelength 
range of 174$-$194 \AA\ is
\begin{equation}
 R^{EIS}_{\lambda}= 10^{a_0+a_1(\lambda-\lambda_0)+a_2(\lambda-\lambda_0)^2}, \label{eqeis} \\ 
\end{equation}
where $\lambda_0$=185 \AA, $a_0$=$-$1.10$\pm$0.03, $a_1$=0.111$\pm$0.003, 
$a_2$=$-$(5.2$\pm$0.6)$\times$10$^{-3}$. Note that the EIS responsivity 
(R$^{EIS}_{\lambda}$) derived above is for 2$^{''}$ slit, while in the case of 
1$^{''}$ slit it should be reduced by a factor of 2.

\subsection{Method 2: EIS calibration by EUNIS-07 LW using insensitive line ratios}
\begin{figure*}
 \epsscale{1.0}
 \plotone{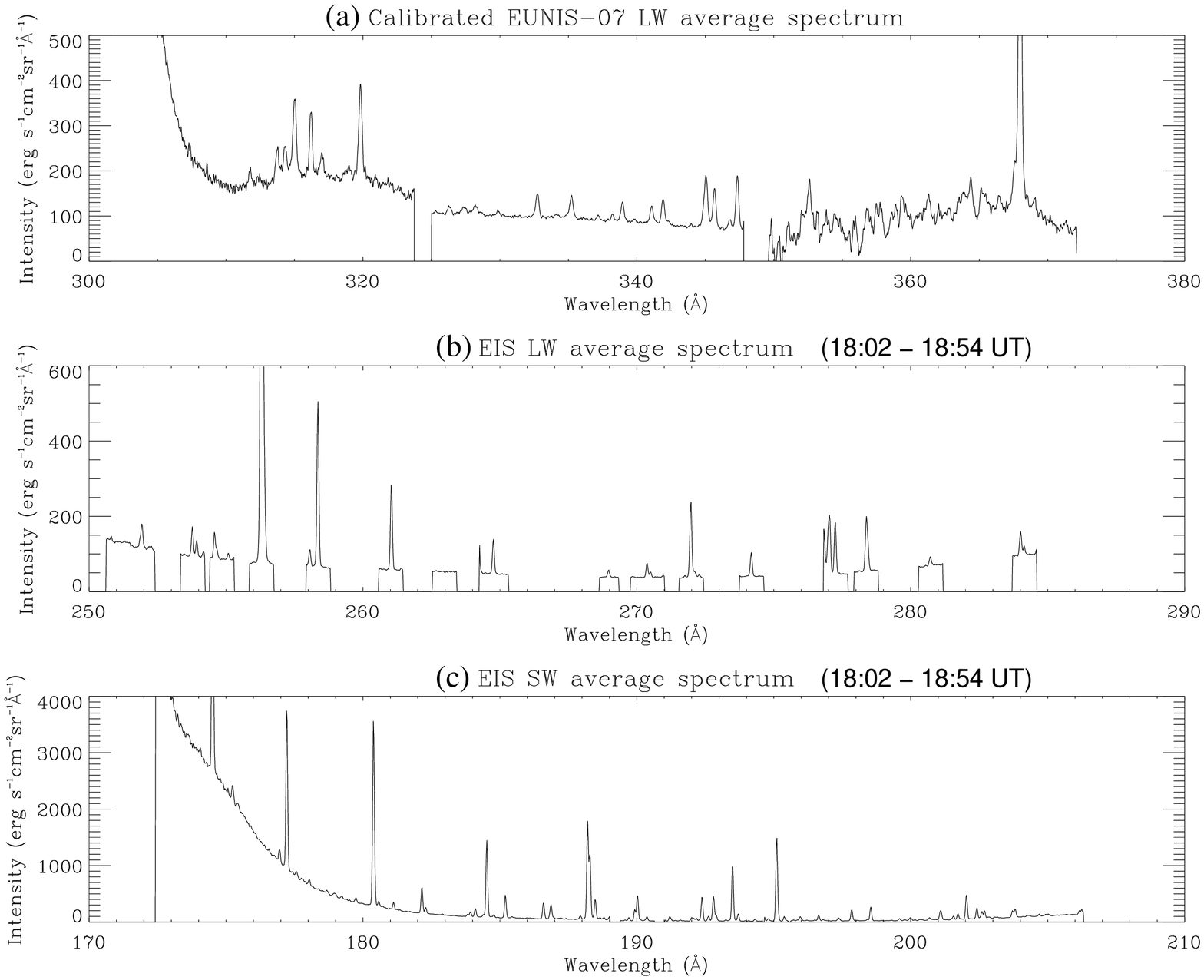}
 \caption{ \label{fglws}
(a) Calibrated EUNIS-07/LW channel spectrum, (b) EIS/LW band spectrum and 
(c) EIS/SW band spectrum, spatially averaged over their common FOV shown 
in Figures~\ref{fgcfv} (a) and~\ref{fgcfv} (b). (b) and (c) are derived with
current EIS calibration (in $SSWIDL$), uncorrected by the factor of 1.22 
derived here. }
 \end{figure*}

\begin{figure}
\epsscale{1.0}
\plotone{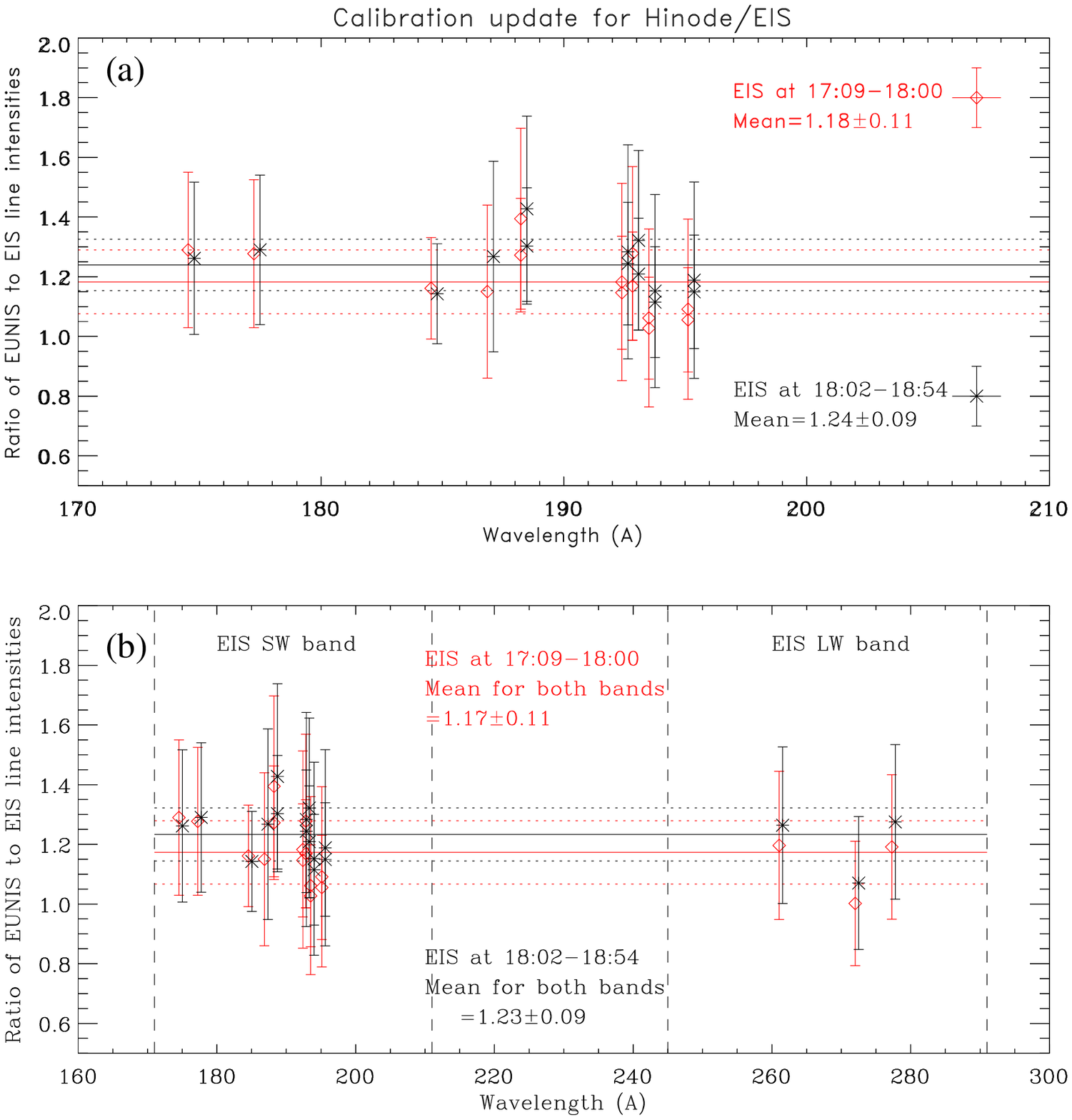}
\caption{ \label{fgrat}
 (a) Plot of the derived-to-measured line intensity ratios for EIS SW band 
in two cases, one for EIS data ({\it diamonds}) observed from 17:09 to 18:00 UT, 
and the other for EIS data ({\it asterisks}) from 18:02 to 18:54 UT. 
(b) The same as (a) but including line intensity ratios for EIS LW band.
 The wavelength of spectral lines in the latter case (EIS data from 18:02
to 18:54 UT) has been increased 
by 0.25 \AA\ in (a) and 0.5 \AA\ in (b) for a clear comparison. In the 
both panels, the thick solid line represents the average ratio and 
the dashed lines their standard deviation.}
\end{figure}

Figure~\ref{fglws} shows the absolutely calibrated spectra for EUNIS-07 LW and EIS 
LW and SW bands, averaged over a common FOV shown in Figures~\ref{fgcfv}(a) 
and~\ref{fgcfv}(b). Table~\ref{tabrat} lists seven groups of temperature- and 
density-insensitive lines from Fe\,{\sc{x}}, Fe\,{\sc{xi}}, Fe\,{\sc{xii}}, and 
Si\,{\sc{x}}. Since the two Fe\,{\sc{xi}} LW lines, 352.66 and 341.11 \AA,
and the two Fe\,{\sc{xii}} LW lines, 352.11 and 364.47 \AA, are insensitive line pairs, 
they were separated forming two groups which include the same SW line members.
Column (3) gives theoretical line intensities normalized to the first member 
(EUNIS LW line) in each group. In Column (4) the absolute intensity of the EUNIS 
LW line was directly measured from the average spectrum, while those of the SW lines 
were derived using the theoretical ratios. Column (5) gives the absolute intensity 
of EIS SW lines measured from the average spectrum. Column (6) are ratios of the 
EUNIS-derived line intensity to the EIS directly measured one, which are shown 
in Figure~\ref{fgrat}. Our measurements from the EIS data set obtained from 18:02 
to 18:54 UT indicate that the EIS responsivities in both bands decreased by a factor of 
1.23$\pm$0.09 compared to the pre-launch ones measured in laboratory, which are 
in good agreement with the result obtained using method 1.

In the case using the EIS data set observed from 17:09 to 18:00 UT prior to the 
EUNIS-07 flight, the obtained line ratios (red symbols shown in Figure~\ref{fgrat})
indicate that the EIS responsivities decreased by a factor of 1.17$\pm$0.11, about 5\%
smaller than that obtained in the former case. This small difference may be caused by
some cooling bright points observed at the southern end of the FOV. By averaging the
results from two EIS data sets, we derived that the responsivities of two EIS bands 
degraded by a factor of 1.2 (or by 17\%) after one year flight although the size 
of the measurement uncertainties is comparable to this decrease.

\section{Discussion and Conclusions}
Using coordinated, cospatial spectroscopic observations of a quiet region of the Sun, 
the lab-calibrated EUNIS-07 LW channel has been directly applied to update the CDS NIS
calibration. We have found that the measured CDS NIS 1 line intensities calibrated with
the standard (version 4) responsivities with the standard long-term corrections are
overall underestimated by a factor of 1.5$\pm$0.6 due to the improper treatment
of the decrease in responsivity, while the EUNIS-07 measurements are in very good 
agreement with the new long-term corrections for NIS 1 derived by \citet{del10}. 
The difference is small (by about 10\%) 
between the He\,{\sc{ii}} 304 intensities measured by EUNIS-07 in first order and 
by CDS NIS 2 in second order with the standard or new long-term corrections.
Therefore the EUNIS-07 and EUNIS-06 have provided the consistent calibration updates 
for CDS NIS 1 in the wavelength range of 301--370 \AA\ and NIS 2 in second order 
at 303--304 \AA\, and confirmed the inflight calibration update by \citet{del10}.
It is worth pointing out both the standard and the new long-term corrections 
for CDS/NIS include a wavelength-dependent decrease in responsivity. A major difference
lies in that the standard correction assumed that the brightest lines would be most 
affected, so applies a correction which is too large for the strongest lines, 
while too small for the weaker lines. The new long-term corrections indicate an overall
decrease in the NIS responsivity at all wavelengths.

The lab-calibrated EUNIS-07 LW channel is also applied to derive the EUNIS-07 SW 
channel calibration using a technique based on density- and temperature-insensitive 
line intensity ratios. The EIS calibration update is performed in two ways. 
One is using the direct calibration transfer of the calibrated EUNIS-07 SW channel. 
The other is using the insensitive line pairs, in which one member was observed 
by EUNIS-07 LW channel and the other by EIS in either LW or SW waveband. The response
curve of EIS SW waveband is derived using the first method, which well coincides 
in shape with the lab-calibrated one prior to launch. Results obtained from the
two methods are in good agreement, and confirm (within the measurement uncertainties) 
the EIS responsivity measured directly before the instrument's launch. The measurements 
also suggest that the responsivity of both EIS wavebands decreased by a factor of 1.2 
(or dropped to 83\% of their prelaunch value) after a year flight (2007 November 6). 
The overall uncertainty due to the slightly temperature- and density-sensitivity of 
theoretical line ratios is less than 10\%. 

Responsivity loss of EUV spectrometers is often measured by monitoring the intensities 
of solar spectral lines from the basal chromospheric-transition region of quiet Sun,
which are known on average not to vary over the cycle \citep[e.g.][]{del10}. 
A recent EIS Science Nugget by Doschek (2011) reported an e-folding time of 6.4 years 
for the EIS instrument's responsivity by measuring the intensity variation of the 
He\,{\sc{ii}} line at 256.32 \AA\ using the synoptic data between January 2007 and 
July 2009. By including quiet Sun data near disk center early in the mission before 
the synoptics started and the synoptic data in 2010, Mariska (2011, private 
communication) updated the measurement of e-fold time to be 5.2 years, which
gives a decrease of responsivity by a factor of 1.2 for the first year's flight, 
very well consistent with that derived in the present work. However,
a preliminary investigation of the EIS radiometric calibration based on CHIANTI Beta v.7
showed that the updated atomic data would reduce the EUNIS-to-EIS line ratios 
by 10\% for Fe\,{\sc{xi}} and Si\,{\sc{x}}, and thus the degradation of EIS
responsivity by a factor of about 1.2  derived here may still be within the uncertainties 
of the EIS prelaunch calibration. In addition, based on the SUMER/EIS joint campaigns 
in 2007 March--April, \citet{mug10, lan10} made the SUMER--EIS relative
intensity calibration and found that the two instruments agreed with each other
within uncertainties.

\begin{deluxetable*}{lllll}
 \tabletypesize{\scriptsize}
 \tablecaption{ Overview of measurements of the He\,{\sc{ii}} 304 \AA\ line 
intensity in quiet regions of the Sun. The values marked with `bl' means blended 
with Si\,{\sc{xi}} 303.4 \AA. The unit of intensities is erg s$^{-1}$cm$^{-2}$sr$^{-1}$.\label{tabhe}}
 \tablewidth{0pt}
 \tablehead{\colhead{Instrument} & \colhead{Obs. Date} & \colhead{Source} & \colhead{Intensity\tablenotemark{a}} & \colhead{Reference} }
\startdata
 EUNIS-07\tablenotemark{b}   & 2007-Nov-06  & Quiet Region & 4960$\pm$496  & This paper \\
 EUNIS-06\tablenotemark{c}   & 2006-Apr-12 & Quiet Region & 4108$\pm$411  & This paper \\
SDO/EVE & 2010 May-Dec             & Full Sun & 5670$\pm$190 (59$\pm$2) bl & This paper \\
Prototype-EVE rocket & 2008-Apr-14 & Full Sun & 5766 (60) bl &  \citet{del11} \\
 CDS NIS		     & 2006-2008   & Full Sun & 5382$\pm$288 (56$\pm$3) & \citet{del11}\\
 CDS NIS\tablenotemark{d}    & 1997-May-15  & Full Sun & 6520$\pm$1304  (68$\pm$14) & \citet{bre00} \\
 CDS NIS\tablenotemark{e}     & 1998-May-1  & Quiet Region & 6856$\pm$132 &  \citet{brk06}\\
 CDS NIS\tablenotemark{e}     & 1996 Mar-Jun & Quiet Region & 7204$\pm$8  &  \citet{war05}\\
 CDS NIS\tablenotemark{f}    & 1997-Nov-18 & AR's Quiet Surroundings &  8520$\pm$144 & This paper\\
 SERTS-97\tablenotemark{g}   & 1997-Nov-18 & AR's Quiet Surroundings  & 8270$\pm$929   & \citet{bro00b} \\
 SERTS-97\tablenotemark{h}   & 1997-Nov-18 & Quiet Region & 7400 (4500--12000)  & This paper \\ 
 SERTS-95\tablenotemark{i}   & 1995-May-15 & Quiet Region & 9510 $\pm$1070 & \citet{bro98b} \\
 SERTS-93\tablenotemark{i}   & 1993-Aug-17 & Quiet Region & 6940$\pm$778 & \citet{bro96} \\
 SERTS-91\tablenotemark{i}   & 1991-May-7  & Quiet Region & 7760$\pm$869  & \citet{bro96} \\
 {\it Skylab} & 1973-Aug & Quiet (cell and network) & 8406 (7527$-$10998) bl & \citet{ver78} \\
 {\it Skylab} & 1973-1974 & Quiet (Center-to-limb) & 7115$-$9321 bl & \citet{man78} \\
 Average for OSO-3,4,6,  & 1967$-$1974 & Full disk & 8413$\pm$1085 bl & \citet{man78}\\
{\it Skylab} and rockets\tablenotemark{j}  & &  & in 7280-11050 bl & 
\enddata
\tablenotetext{a}{Those values for Full Sun (in lines 3-6) are converted 
from the irradiance in bracket which has units of 10$^8$ photons~s$^{-1}$~cm$^{-2}$.}
\tablenotetext{b}{Average intensity for the common FOV between LW and SW channels.}
\tablenotetext{c}{Average intensity for a very quiet area near an active region 
        with the absence of Fe\,{\sc{xiv}} line emission.}
\tablenotetext{d}{Recalibration of SOHO/CDS NIS with the EUV Grating Spectrograph 
  (EGS) on the NASA/LASP rocket. }
\tablenotetext{e}{The values in \citet{war05,brk06} are doubled to correct the calibration
(Warren \& Brooks 2011, private communication). }
\tablenotetext{f}{Average of two measurements for observations at 18:40 and 19:49 UT.}
\tablenotetext{g}{Average intensity for a quiet area near an active region 
observed at SERTS pointing position 2 \citep{swa99}. }
\tablenotetext{h}{Average intensity for the quiet regions observed at SERTS 
  pointing position 1 \citep{swa99}. }
\tablenotetext{i}{Calibrations were made based on the `standard' quiet Sun 
intensity given by \citet{man78}.}
\tablenotetext{j}{Average of 9 measurements taken from Table 2 of \citet{man78}.}
\end{deluxetable*}

The quiet Sun emission of the He\,{\sc{ii}} 304 \AA\ line has been known for 
a long time to remain fairly constant. Early observations from Rocket, Skylab and 
OSO in 1960-70s give typical intensities of 7000$-$9000 erg s$^{-1}$cm$^{-2}$sr$^{-1}$  
for the He\,{\sc{ii}} + Si\,{\sc{xi}} blend (see Table~\ref{tabhe} and \citet{del11}
who summarized almost all historical records). 
The center-to-limb curve for quiet Sun He\,{\sc{ii}} + Si\,{\sc{xi}} 
intensity of such values reported by \citet{man78} was used as a natural
``standard'' light source for absolute radiometric calibrations of SERTS-89, SERTS-91, SERTS-93 
and SERTS-95 \citep{tho94, bro96, bro98b}. SERTS-97 was directly calibrated at RAL 
in the same facility used to characterize the CDS and EUNIS instruments, and provided 
the calibration update for CDS NIS-1 first-order and NIS-2 second-order lines. SERTS-97 pointed at
two locations during its observation \citep{swa99}. The slit position in pointing 1 
was in the quiet region with average He\,{\sc{ii}} intensity of 7400 
erg~s$^{-1}$cm$^{-2}$sr$^{-1}$, while it covered the ``quiet surroundings" of an AR 
in pointing 2 with average He\,{\sc{ii}} intensity 8270 erg~s$^{-1}$cm$^{-2}$sr$^{-1}$
\citep{bro00b}.  The quiet-Sun He\,{\sc{ii}} radiance measured by SERTS-97 is 
consistent with the value measured near the disk center by \citet{man78} from Skylab,
and is also well consistent with those measured by CDS NIS 2 during 1996--1998 
\citep[e.g.][]{bre00, war05, brk06}. 

However, we find that the quiet Sun He\,{\sc{ii}} radiances measured by EUNIS-07 on 
the order of 5000 erg s$^{-1}$cm$^{-2}$sr$^{-1}$, which is about a factor of 1.4 smaller 
than that measured by SERTS-97 and about a factor of 1.6 smaller than the old 
measurements in 1960--70s (see Table~\ref{tabhe}). Note that the old measurements 
include the blending emission from Si\,{\sc{xi}} line, but the blending typically 
contributes less than 10\% of the total intensity of these two lines in the quiet Sun, 
so cannot explain the big difference from the EUNIS-07 measurement. 
Based on the long-term monitoring of CDS NIS 
irradiances from 1998 to 2010, \citet{del11} suggested that the Skylab 
values of the He\,{\sc{ii}} 304 \AA\ line in the quiet Sun were overestimated, and 
so did for those by SERTS-89, SERTS-91, SERTS-93 and SERTS-95 whose absolute radiometric
calibration was based on the Skylab measurements. Whereas the large 
difference between the SERTS-97 and EUNIS-07 measurements which were both based on 
the direct lab-calibration
suggests that the large dispersion of measurements from different instruments may be 
partially ascribed to true variations depending on observed locations if the average is 
limited in a small FOV, and on observing times (e.g. different phases of a solar cycle, 
or different solar cycles). For example, the intensity of the He\,{\sc{ii}} 304 \AA\ 
line in ``quiet surroundings" of an active region may be significant larger than 
that in the very quiet region. EUNIS-06 showed that the ``quiet surroundings" have 
the values of 6000$-$8000 erg~s$^{-1}$cm$^{-2}$sr$^{-1}$, while the very quiet region 
considered where Fe\,{\sc{xiv}} emission is nearly absent has the value of 
about 4100 erg~s$^{-1}$cm$^{-2}$sr$^{-1}$. The SERTS-97 observation at pointing 1 showed
variations of He\,{\sc{ii}} intensities in ``quiet region" from 4500 to 12000 
erg~s$^{-1}$cm$^{-2}$sr$^{-1}$ over the 353$^{''}$-long slit.
\citet{del11} showed that the He\,{\sc{ii}} irradiance measured during this solar 
minimum (2006--2008) is about a factor of 1.3 smaller than that measured in 1998 
at the beginning of the last solar cycle. The dispersion of 
He\,{\sc{ii}} intensities shown in Table~\ref{tabhe} suggests that
one should be cautious about using the quiet-Sun He\,{\sc{ii}} line as a standard
light source. 

In addition, we may verify the EUNIS-07 radiometric calibration by comparing the measured
quiet-Sun He\,{\sc{ii}} 304 \AA\ intensity with the radiance converted from irradiance 
measurements (the full-disk flux measured at Earth from this line or a narrow waveband
centered at 304 \AA) during the solar minimum. Since the He\,{\sc{ii}} lines have 
negligible limb-brightening and off-limb contribution \citep{del11}, the conversion 
for the quiet Sun can be simply made by the relation \citep{war98}
\begin{equation}
 F_{qs}=\frac{\pi R^2_{\sun}}{R^2} I_{qs},
\end{equation}
where $F_{qs}$ is the irradiance, $R_{\sun}$ is the solar radius, $R$ is the 
Earth-Sun distance, and $I_{qs}$ is the
intensity at disk-center. When the irradiance uses the unit of photons s$^{-1}$ cm$^{-2}$
and the intensity is in erg~s$^{-1}$cm$^{-2}$sr$^{-1}$, the conversion coefficient
between $F_{qs}$ and $I_{qs}$ at 304 \AA\ is 1.04$\times$10$^{6}$ sr~photons~erg$^{-1}$.  
From the quiet-Sun intensity of 4960 erg~s$^{-1}$cm$^{-2}$sr$^{-1}$ by EUNIS-07, it derives 
irradiance to be $52\times10^8$ photons s$^{-1}$ cm$^{-2}$, which is well consistent 
with recent measurements ((56$\pm3$)$\times10^8$ photons s$^{-1}$ cm$^{-2}$) 
by CDS NIS-2 during the solar minimum from 2006 to 2008 \citep{del11}, and those 
(about 60$\times10^8$ photons s$^{-1}$ cm$^{-2}$ including the blended Si\,{\sc{xi}} line)
by prototype-EVE rocket in April 2008 \citep{del11} and by SDO/EVE over May-December 2010 
(see Table~\ref{tabhe}). These comparisons verify the good radiometric
calibration for EUNIS-07.   

Finally, we may check the absolute intensity of the quiet-Sun He\,{\sc{ii}} 304\AA\ 
line from EUNIS-07 using the theoretical line ratio and coordinated, cospatial EIS
measurements. The CHIANTI (ver.6.0) shows that the He\,{\sc{ii}} 256 \AA/304 \AA\ 
ratio in optically thin condition is somewhat density sensitive, with a minimum 
value of 0.036 at log$N_e$ = 8.0, and a maximum value of 0.117 at log$N_e$ = 11.0, 
but a most likely value of 0.051$\pm$0.011 for 9.0$\leq$log$N_e$$\leq$10.0. In comparison,
\citet{jor75} gave the observed ratio of this line pair to be 0.052, and 
\citet{bro98b} measured the ratio of He\,{\sc{ii}}+Si\,{\sc{x}} 256\AA\ to 
He\,{\sc{ii}} 304\AA\ to be 0.046$\pm$0.008 for an active
region. These measurements agree with the predicted value within uncertainties. 
Taking the theoretical ratio from CHIANTI and the measured EIS intensity of 
the He\,{\sc{ii}} 256 \AA\ line (with calibration correction by EUNIS-07, ie.
scaled by a factor of 1.2), we derived the intensity of He\,{\sc{ii}} 304\AA\ to be 
4810$\pm$1040 for the EIS 17:09$-$18:00 UT raster and 4630$\pm$1000 for the EIS 
18:02$-$18:54 UT raster. Despite the He\,{\sc{ii}} lines being possibly optically 
thick, the quiet-Sun He\,{\sc{ii}} 304 \AA\ intensity derived from EIS is well 
consistent with that directly measured by EUNIS-07, confirming an agreement 
between the EUNIS-07 and EIS cross-calibration.

In the next flight (expected in 2011 November) EUNIS will focus on the study 
of thermal structure of coronal loops and coordinated observations with SDO/AIA. 
Based on SDO images, two slit locations
and rotational alignments will be chosen before flight to sample both active regions
and quieter areas. EUNIS-11 will continue to provide the EIS cross-calibration as done with
EUNIS-07 using the insensitive line ratio technique. Active region spectra will
allow applications of more line pairs such as from Fe\,{\sc{xiii}}$-$Fe\,{\sc{xvi}} to
provide a good coverage for the EIS LW band. 

In conclusion, EUNIS-07 has provided important underflight calibration updates for
CDS/NIS and Hinode/EIS. The results by EUNIS-07 well support the recent measurements
of the long-term correction for the CDS and EIS instrument responsivities 
obtained by the inflight monitoring of the quiet-Sun intensity 
of the chromospheric-transition region lines. The absolute value of the quiet-Sun
He\,{\sc{ii}} 304 \AA\  intensity measured by EUNIS-07 is well consistent with 
the radiance measured by CDS NIS in quiet regions near the disk center and 
the solar minimum irradiance obtained by CDS NIS and SDO/EVE recently, 
but it is about a factor of 1.4--1.6 smaller than the previous values from SERTS-97 
and Skylab, reflecting possible real variations with locations and times.

\acknowledgments
The EUNIS program is supported by the NASA Heliophysics Division through its Low Cost 
Access to Space Program in Solar and Heliospheric Physics. TW is grateful to 
Drs. William T. Thompson, John Mariska and Vincenzo Andretta for their valuable comments. 
The work of TW was supported by NASA grants NNX10AN10G and NNX08AE44G. The work of PRY was performed 
under contract with the Naval Research Laboratory and was funded by NASA.
GDZ acknowledges support from STFC (UK) via the Advanced Fellowship programme.
Radiometric calibration of the EUNIS-06 instrument was made
possible by financial contributions and technical support from both the Rutherford-Appleton 
Laboratory in England and the Physikalisch-Technische Bundesanstalt in Germany, 
for which we are very grateful. CHIANTI is a collaborative project involving 
the Universities of Cambridge (UK), George Mason and Michigan (USA).

\end{document}